\documentclass[aps,groupedaddress, twocolumn,  nofootinbib,prd]{revtex4-1}
\usepackage{latexsym}
\usepackage{amsmath}
\usepackage{amsfonts}
\usepackage{amsbsy}
\usepackage{amssymb}
\usepackage{epsfig}
\usepackage{mathrsfs}
\usepackage{epsfig}
\usepackage{bbm}
\usepackage{color}

\begin{document}

\title{{\bf Renormalization of symmetry restricted spin foam models with curvature in the asymptotic regime}}

\author{Benjamin Bahr}
\email[]{benjamin.bahr@desy.de}
\author{Giovanni Rabuffo}
\email[]{giovanni.rabuffo@desy.de}
\affiliation{II. Institute for Theoretical Physics\\University of Hamburg\\ Luruper Chaussee 149\\22761 Hamburg\\Germany}
\author{Sebastian Steinhaus}
\email[]{ssteinhaus@perimeterinstitute.ca}
\affiliation{Perimeter Institute for Theoretical Physics\\31 Caroline St N\\N2L 2Y5 Waterloo, ON\\Canada}


\date{\today}

\begin{abstract}
We study the renormalization group flow of the Euclidean Engle-Pereira-Rovelli-Livine and Freidel-Krasnov (EPRL-FK) spin foam model in the large-$j$-limit. The vertex amplitude is deformed to include a cosmological constant term. The state sum is reduced to describe a foliated spacetime whose spatial slices are flat, isotropic and homogeneous. 
The model admits a non-vanishing extrinsic curvature whereas the scale factor can expand or contract at successive time steps.  \\
The reduction of degrees of freedom allows a numerical evaluation of certain geometric observables on coarser and finer discretizations.
Their comparison defines the renormalization group (RG) flow of the model in the parameters $(\alpha,\Lambda,G)$.
We first consider the projection of the RG flow along the $\alpha$ direction, which shows a UV-attractive fixed point. Then, we extend our analysis to two- and three-dimensional parameter spaces. Most notably, we find the indications of a fixed point in the $(\alpha,\Lambda,G)$ space showing one repulsive and two attractive directions.

\end{abstract}


\maketitle
\section{Motivation}

Loop Quantum Gravity (LQG) and its covariant cousins, Spin Foam Models (SFM), are among the most promising approaches to defining a theory of quantum gravity \cite{thomasbook, cbook, Perez:2012wv, cfbook}. These models have advanced significantly over the last two decades, and possess a solid theoretical foundation. 

However, any approach to quantum gravity must eventually walk the long and hard road towards making contact with observations and attempt to make predictions. This is largely uncharted territory for  LQG and SFM, in the sense that they are not yet in a form in which they can be used to produce reliable numbers which can be compared to experiment. A notable exception to this is the cosmological version of Loop Gravity, Loop Quantum Cosmology (LQC) \cite{Ashtekar:2011ni}. Also, some SFM are used to estimate the lifetime of black holes, and attempt to search for observable signals of their decay into white holes \cite{Christodoulou:2016vny}.

In general, however, it is quite difficult to use LQG and SFM to produce reliable numerical predictions. One of the main reasons for this is the fact that in their background-independent formulation, LQG and SFM are far away from the continuum physics of either General Relativity or Quantum Field Theory. A direct comparison is therefore quite difficult, and computations of, say, quantum gravity corrections to known processes, are quite hard to do. Therefore, the problem is intimately tied to the question of their continuum limit. 


Both LQG and SFM describe (respectively canonical and covariant) dynamics for the microscopic degrees of freedom of space-time. Gravity, as an interactive field theory, can be expected to have a non-trivial RG flow, and the effective dynamics for macroscopic degrees of freedom, which is what we measure and describe by GR, can differ radically from the microscopic theory\footnote{A striking example for this is the description of cosmological space-time as condensate of building blocks of space, as can be described in the group field theory (GFT) approach \cite{Oriti:2007qd, Gielen:2013naa}.}. To understand the continuum limit of these theories, one therefore needs to compute their RG flow, i.e.~the way in which the theory effectively changes among different scales \cite{Wilson:1973jj}.

Both LQG and SFM are constructed with background-independence in mind, in order to incorporate Einstein's principle of general covariance. This prevents the direct application of well-established RG methods, since these contain a notion of scale which relies on a background metric\footnote{In lattice gauge theory, one e.g.~works on fixed lattices with a lattice length $a$, which depends on the background geometry.}. In recent years, however, there have been significant advances in understanding the notion of  background-independent renormalization group flow, in which the notions of coarse graining, scale, and RG flow are generalized to the setting of LQG  and SFM \cite{Dittrich:2011zh, Bahr:2011aa, Bahr:2012qj,  Dittrich:2012jq, Dittrich:2014ala, Bahr:2014qza, Dittrich:2016tys, Bahr:2016hwc, Bahr:2017klw, Lang:2017beo, Lang:2017yxi, Lang:2017oed, Lang:2017xrb}. 

These methods bear close resemblance to those developed for tensor network renormalization (TNR \cite{levin,guwen,vidal-evenbly}), and have been successfully applied to SFM in 2D and 3D \cite{Dittrich:2011zh,Dittrich:2013bza,Dittrich:2016tys}. The development of further approximation techniques have allowed to also apply these methods to SFM in a truncated setting. 

The RG flow of 4D models yielded quite interesting results. In particular, it was observed that the RG flow of the so-called EPRL-FK Spin Foam model \cite{Engle:2007wy, Freidel:2007py, Kaminski:2009fm} appears to possess a non-trivial UV fixed point \cite{Bahr:2016hwc, Bahr:2017klw}. While promising, it is unclear how much these results depend on the truncations of the EPRL-FK model, which were used. Therefore, in this article, we are going to relax some of these truncations, in order to investigate the EPRL-FK model in 4D in more detail. 

These relaxations constitute a significant extension of the analysis in \cite{Bahr:2016hwc}, in that we do not only consider the path integral over a specific diffeomorphism orbit, but sum over configurations with different curvature, which could not be regarded as diffeomorphically equivalent. We therefore expect this to tell us much more about the whole path integral than previous investigations.

The outline of the paper is as follows:\\

First we describe the general setup of our article in chapter \ref{Sec:Setup}. In section \ref{Sec:SFM} we remind the reader of the definition of the EPRL-FK model, which is used in our analysis. In section \ref{Sec:BIR} we review the notion of coarse graining, in the background-independent context for SFM. We discuss the approximations and truncations used in this and previous articles in section \ref{Sec:Approximations}. In section \ref{Sec:Numerics} we give an overview of the numerical methods employed in our analysis.

In section \ref{Sec:IsotemporalRGFlow} we consider the RG flow of the 4D EPRL model, in various gauges and truncations. We consistently find a similar fixed point to the one in \cite{Bahr:2016hwc, Bahr:2017klw}, confirming and extending the results from these earlier works. In section  \ref{Sec:ExpandingContracting} we additionally consider the dynamics of the symmetry-restricted model numerically, in order to gain some insight into the behaviour of the model. This also lends some interpretation to the RG flow analysis. In \ref{Sec:FreeTheory}, we consider the asymptotic limit of the free theory, which is suspected to be a Gaussian fixed point of the RG flow. We sum up our results in \ref{Sec:Summary}.


\section{General Setup}\label{Sec:Setup}

Spin Foam models describe the dynamics of quantum gravity by assigning transition amplitudes to Loop Quantum Gravity boundary states. There exist various different versions of spin foam models \cite{Barrett:1997gw, Engle:2007wy, Freidel:2007py, CubulationSpinFoamThiemann2008, Baratin:2011hp}
for both Riemannian and Lorentzian signature. In this article, we will focus on the so-called EPRL-FK model \cite{ Engle:2007wy, Freidel:2007py}. For simplicity, we consider Riemannian signature, so that the local gauge group is $SU(2)\times SU(2)\simeq \text{Spin}(4)$. 

\begin{figure}[h]
	\centering
	\includegraphics[scale=0.5]{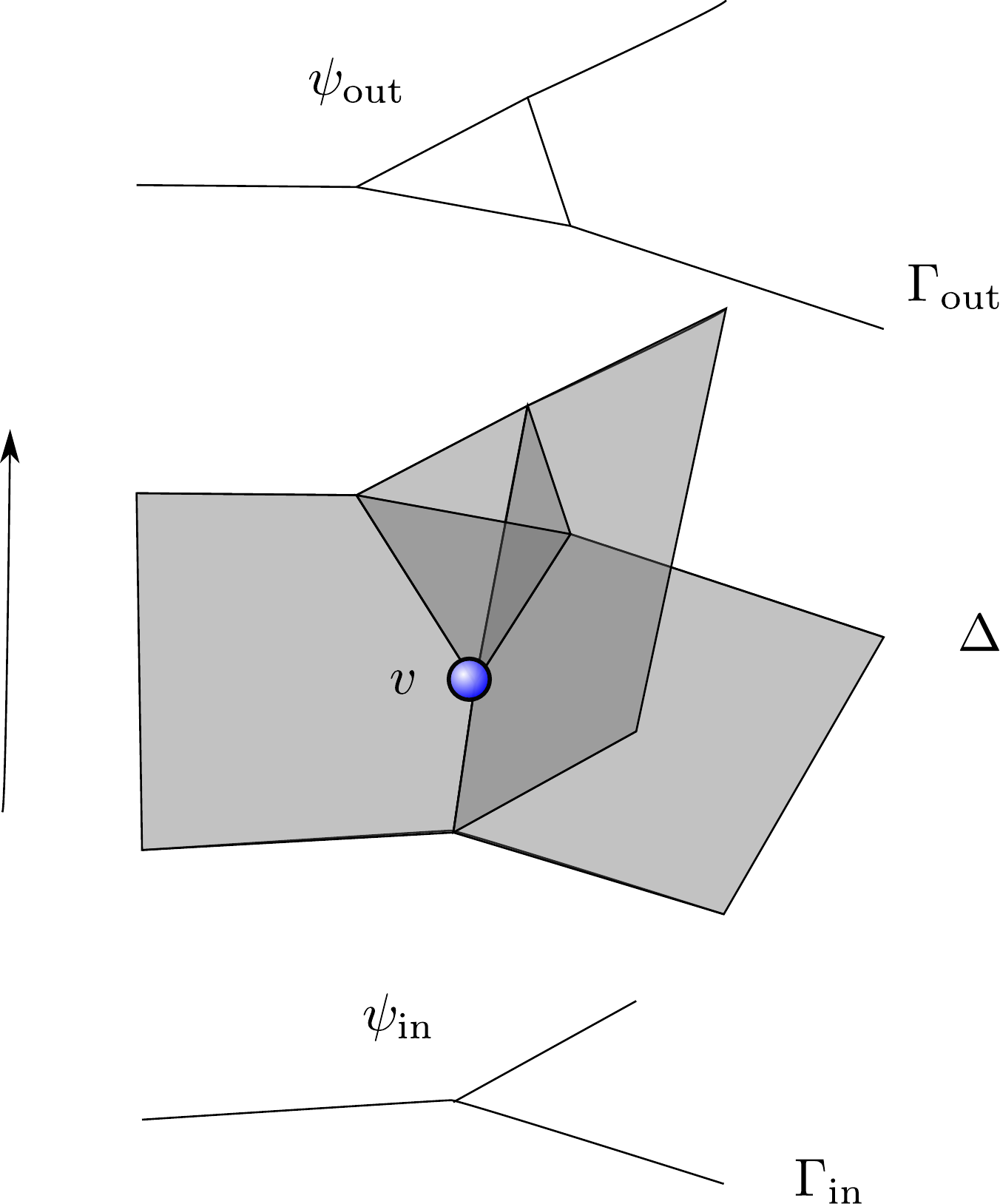}
	\caption{
		\small A spin foam $\Delta$ describes the history of a spin network boundary graph. The interaction vertices $v$ are where the topology of the graph can change. }
	\label{Fig:Figure_01}
\end{figure}

The boundary states of SFM are given by a generalization of Penrose's spin network functions to general graphs $\Gamma$. A transition is described by a history of a graph, called a spin foam. Since a graph consist of $1$-dimensional parts (links) and 0-dimensional parts (nodes), the elements of a spin foam $\Delta$ are 2-dimensional (faces, the history of a link), 1-dimensional (edges, the history of a node), and 0-dimensional (vertices, where the topology of a graph can change). See figure \ref{Fig:Figure_01}. These are often taken to be the dual 2-complex to a polyhedral decomposition of space-time, but they can be more general 2-complexes \cite{Kaminski:2009fm}.

A spin foam model is specified by an assignment of amplitudes to boundary graphs. 

\subsection{The EPRL spin foam model}\label{Sec:SFM}

We consider a general spin foam vertex, for the Riemannian signature EPRL-FK model, with Barbero-Immirzi paramter $\gamma\in(0, 1)$. The associated amplitude is a linear map on the boundary Hilbert space. A state in that Hilbert space is given by boundary data, which is completely described by a directed graph $\Gamma\subset S^3$ embedded into a three-sphere.

A boundary geometry on $\Gamma$ is given by a collection of spins $j_L\in\frac{1}{2}\mathbb{N}$ associated to the links $L\in \text{Links}(\Gamma)$ of $\Gamma$, and a collection of $3d$ unit vectors $\vec{n}_{NL}$ associated to pairs of nodes $N\in \text{Nodes}(\Gamma)$ of the graph, and links $L$ which are connected to $N$. For all $L\supset N$, the corresponding unit vectors are chosen such that they satisfy
\begin{eqnarray}\label{Eq:Closure}
	G_N\;:=\;\sum_{L\supset N}j_L \vec{n}_{NL}\;=\;0.
\end{eqnarray}

\noindent For Riemannian signature, the local gauge group is $\text{Spin}(4)$. We use the Hodge duality in four dimensions, under which its Lie algebra decomposes into $\mathfrak{spin}(4)\simeq \mathfrak{su}(2)\oplus\mathfrak{su}(2)$, two commuting $SU(2)$-subalgebras, which are the eigenspaces under the Hodge $*$ for eigenvalues $\pm 1$. Consequently, one has the group isomorphism $\text{Spin}(4)\simeq SU(2)\times SU(2)$, and an irreducible representation of $\text{Spin}(4)$ can therefore be depicted as pair $(j^+,j^-)$ of half-integers.

The vertex amplitude $\mathcal{A}_v$ is constructed in the following way: Define\footnote{With the definition (\ref{Eq:SpinsSimplicity}), one has to demand that all three $j_L, j_L^\pm$ are half-integers, which puts severe restrictions on the Barbero-Immirzi parameter $\gamma$. This is a pathology of the Riemannian model, which does not occur in the Lorentzian context.}
\begin{eqnarray}\label{Eq:SpinsSimplicity}
	j^\pm_L\;:=\;\frac{|1\pm \gamma|}{2}j_L,
\end{eqnarray}

\noindent and construct the boosted \emph{Livine-Speziale-intertwiners} \cite{Livine:2007vk, Bianchi:2010gc}
\begin{eqnarray}\label{Eq:BoostedIntertwiner}
	\iota_N^\pm\;:=\;\mathcal{P}\left[\bigotimes_{L\leftarrow N} \beta_L|j_L,\,\vec{n}_{NL}\rangle
	\otimes\bigotimes_{L\rightarrow N}\langle j_L,\,-\vec{n}_{NL}|\beta_L^\dag \right].
\end{eqnarray}

\noindent The $SU(2)\times SU(2)$-intertwiner $\iota_N^\pm=(\iota_N^+,\iota_N^-)$ factorises for $\gamma<1$. Here the coherent states for $\vec{n}\in S^2$ are given by
\begin{eqnarray}
	|j,\,\vec{n}\rangle\;:=\;D_j(g_{\vec{n}})|j\;j\rangle,
\end{eqnarray}

\noindent i.e.~the action on the highest weight vector with a group element $g_{\vec{n}}$, which is such that $g_{\vec{n}}e_z=\vec{n}$, with $e_z$ being the unit vector in $z$-direction.\footnote{Note that, given $n\in S^2$, the corresponding $g_{\vec{n}}$ is only defined uniquely up to a $U(1)\subset SU(2)$-subgroup. Different choices amount to different states $|j,\,\vec{n}\rangle$, which differ by a complex phase. For one vertex amplitude, this phase is not important, while for larger triangulations, the relative phases of these states in neighboring vertices have to be taken care of, since they encode the $4d$ curvature. } Note that $D_j(g)$ is the representation matrix of $g\in SU(2)$ for the irreducible representation $j$.

The map
\begin{eqnarray}
	\beta_L : V_{j_L}\;\longrightarrow\;V_{j_L^+}\otimes V_{j_L^-} \; ,
\end{eqnarray}

\noindent is the isometric embedding of $j_L$ into the highest weight subspace of the Clebsh-Gordon decomposition of
\begin{eqnarray}
	V_{j_L^+}\otimes V_{j_L^-}\simeq V_{|j_L^+-j_L^-|}\oplus \cdots V_{j_L^++j_L^-},
\end{eqnarray}

\noindent and $\mathcal{P}:\mathcal{H}\to\text{Inv}_{SU(2)\times SU(2)}(\mathcal{H})$ with
\begin{eqnarray}
	\mathcal{H}\;:=\;\left(\bigotimes_{L\leftarrow N}V_{j_L^+}\otimes V_{j_L^-}\right)\otimes \left(\bigotimes_{L\rightarrow N}V_{j_L^+}^\dag\otimes V_{j_L^-}^\dag\right),
\end{eqnarray}

\noindent is the projector onto the invariant subspace of the Hilbert space $\mathcal{H}$.

As a result of this definition, the tensor product of all boosted Livine-Speziale intertwiners (\ref{Eq:BoostedIntertwiner}) is an endomorphism on the tensor product of all representation spaces over the links, i.e.~
\begin{eqnarray}
	\bigotimes_N\;\iota_N^\pm\;:\;\bigotimes_L\left(V_{j_L^+}\otimes V_{j_L^-}\right)\;\longrightarrow\;\bigotimes_L\left(V_{j_L^+}\otimes V_{j_L^-}\right).
\end{eqnarray}

\noindent The vertex amplitude $\mathcal{A}_v$ is defined as the trace of this map, i.e.~
\begin{eqnarray}\label{Eq:VertexTrace}
	\mathcal{A}_v\;:=\;\text{tr}\left(\bigotimes_N\;\iota_N^\pm\right)\;=\;\mathcal{A}^+_v\mathcal{A}^-_v.
\end{eqnarray}

\noindent The spin foam state sum $Z$ for a larger $2$-complex is defined by summing over several different products of vertex amplitudes. A $2$-complex is regarded as the history of a spin network \cite{Reisenberger:1996pu}, and consists of vertices $v$, edges $e$ and faces $f$. For every vertex, we denote the vertex graph $\Gamma(v)$ to be the one which has a node for every edge touching $v$, and a link between two nodes whenever two such edges being in the boundary of the same face (this is the boundary graph for the spin foam consisting only of the neighbourhood of $v$). Any assignment of spins $j_f^\pm$ to faces, which satisfy (\ref{Eq:SpinsSimplicity}), and corresponding intertwiners $\iota_e^\pm$ to edges, induces a boundary state to every vertex graph, and we denote the vertex amplitudes $\mathcal{A}_v$ to be the corresponding traces (\ref{Eq:VertexTrace}).

For a $2$-complex without boundary, the formal spin foam sum is then defined as
\begin{eqnarray}\label{Eq:GeneralSetup_EPRL}
	Z\;=\;\sum_{j_f^\pm,\iota_e^\pm}\prod_{f}\mathcal{A}_f\prod_e\mathcal{A}_e\prod_v\mathcal{A}_v,
\end{eqnarray}

\noindent where the face- and edge amplitudes are defined by
\begin{eqnarray}\label{Eq:GeneralSetup_FaceAmplitude}
	\mathcal{A}_f\;&=&\;\Big((2j^+_f+1)(2j_f^-+1)\Big)^\alpha \; , \\[5pt]\label{Eq:GeneralSetup_EdgeAmplitude}
	\mathcal{A}_e\;&=&\;\|\iota_e\|^{-2} \; .
\end{eqnarray}

\noindent The parameter $\alpha$ plays the role of a coupling constant, in that it is a free parameter in the path integral measure.  The sum effectively can be expressed as a sum over $SU(2)$-spins $j_f$ and $SU(2)$-intertwiners $\iota_e$, which we will do from now on. 

All three amplitude types are local functions of the spins and intertwiners. For $2$-complexes with boundary, the amplitudes for edges and faces meeting the boundary have to be altered. In essence, we choose the boundary amplitudes 
\begin{eqnarray}
	\mathcal{B}_e=\big(\mathcal{A}_e\big)^\frac{1}{2},\quad \mathcal{B}_f=\big(\mathcal{A}_f\big)^\frac{1}{n_f},
\end{eqnarray}
\noindent where $n_f$ is the number of partial faces that are glued together to form the whole face. See \cite{Bahr:2010bs, Bahr:2012qj, Dittrich:2012er} for details. Since we work with hypercubic lattices, we will use $n_f=4$ throughout this article. 

In this case, due to the regularity of the lattice, one can repackage the sum, to write
\begin{eqnarray}\label{Eq:GeneralSetup_StateSum}
	Z\;=\;\sum_{j_f,\iota_e}\prod_v \hat{\mathcal{A}}_v
\end{eqnarray}

\noindent with 
\begin{eqnarray}\label{Eq:RepackagedSpinFoamSum}
	\hat{\mathcal{A}}_v\;:=\;\prod_{f\supset v}\mathcal{A}_f^{\frac{1}{4}}\prod_{e\supset v}\mathcal{A}_e^{\frac{1}{2}}\mathcal{A}_v.
\end{eqnarray}

\noindent The spin foam sum can be written in terms of boundary amplitudes in the following way: For each vertex $v$ and configuration $j_f$, $\iota_e$, a $SU(2)$-spin network function $\psi_{\Gamma(v),j_f,\iota_e}$ is induced on the corresponding boundary graph $\Gamma(v)$. The boundary amplitude $\mathcal{A}_\Gamma(v)$ is then an operator on $\mathcal{H}_{\Gamma(v)}$, which is defined by
\begin{eqnarray}\label{Eq:GeneralSetup_LinearMap}
	\mathcal{A}_{\Gamma(v)}(\psi_{\Gamma(v),j_f,\iota_e})\;:=\;\hat{\mathcal{A}}_v.
\end{eqnarray}

\subsection{Background-independent renormalization}\label{Sec:BIR}

Renormalization in this article is understood in the Wilsonian sense \cite{Wilson:1973jj}. A theory with infinitely many degrees of freedom is usually formulated in terms of effective theories on only part of those degrees of freedom, e.g.~by introducing a lattice or a momentum cut-off, i.e.~a scale. This effective theory then depends on the scale, usually by scale-dependent parameters called coupling constants.

In the background-independent setting of spin foam models, lengths or energy are encoded in the variables, not in any background structure. This prevents the use of e.g.~a fixed lattice spacing to characterize the scale, and requires one to generalize the well-established renormalization group methods from e.g.~lattice gauge theory. This has been achieved in recent years \cite{Oeckl:2002ia, Bahr:2011aa, Bahr:2012qj,Dittrich:2012jq, Dittrich:2013xwa, Bahr:2014qza, Dittrich:2014ala}, and has led to a version of the RG flow in which space-time discretization itself is regarded as scale.\footnote{Note that on a fixed geometry, refinement of the lattice is equivalent to shrinking of lattice length, while in the background-independent setting of spin foam models, only the former can be defined, since the lattice spacing is a variable to be summed over in the path integral.} Hence the scale is taken to be the 2-complex $\Delta$ itself. The regularization is understood as restricting the theory to only finitely many holonomies, i.e.~those which are associated to $\Delta$.

It should be noted that this is a deviation from the usual way of renormalization, which associates the scale of the flow with a value of maximal energy or minimal length, which are introduced as cutoff. This deviation is an important consequence of this particular way of background-independence of the model. For a given background-geometry, a finer and finer regular lattice leads to an ever-decreasing value of lattice constant. However, in our case these two notions are disentangled, in that on coarse or fine lattices both small and large spins occur. The reason is that on both lattices the geometry is not fixed, but rather the path integral sums over all of them. Hence, the notion of refinement of lattices is the only one that remains in this particular way of dealing with the sum over geometries.

As a consequence, notions of UV and IR limit are not associated to e.g.~small and large spins, but rather to fine and coarse lattices. 

On a technical level, this makes the RG flow procedure very similar to those employed e.g.~in tensor network renormalization \cite{guwen, levin}, see also e.g.~\cite{Dittrich:2011zh}. Of course, this begs the question how the results of this article compare to ones obtained in similar approaches, such as quantum Regge calculus, see e.g.~\cite{Hamber:1990dw,Hamber:1992jh,Hamber:2005dw, Hamber:2006sv}, or causal dynamical triangulations \cite{Ambjorn:2014gsa}. We refer to the discussion in \cite{Bahr:2015gxa}, although this is still an open question at this point.

In the background-independent framework for renormalization we employ, the spin foam sum (\ref{Eq:GeneralSetup_StateSum}), is understood as an effective theory for the available degrees of freedom provided by the $2$-complex. It can be seen as the result of integrating out all of the finer degrees of freedom, which are below the lattice resolution. The lattice itself, then, can be regarded as the result of successive coarse graining of a much finer lattice. 

The question, then, is how the theories on different lattices, i.e.~on different scales, are related. Mathematically, the amplitudes are given in terms of linear maps on the boundary graphs $\Gamma$ of vertices (\ref{Eq:GeneralSetup_LinearMap}). However, several of them together can be made to a linear map onto a larger lattice, with refined boundary graph $\Gamma'$. This allows to rewrite the RG flow of bulk lattices into equations for boundary amplitudes. See also \cite{Dittrich:2014ala}.

\begin{figure*}
	\includegraphics[scale=0.15]{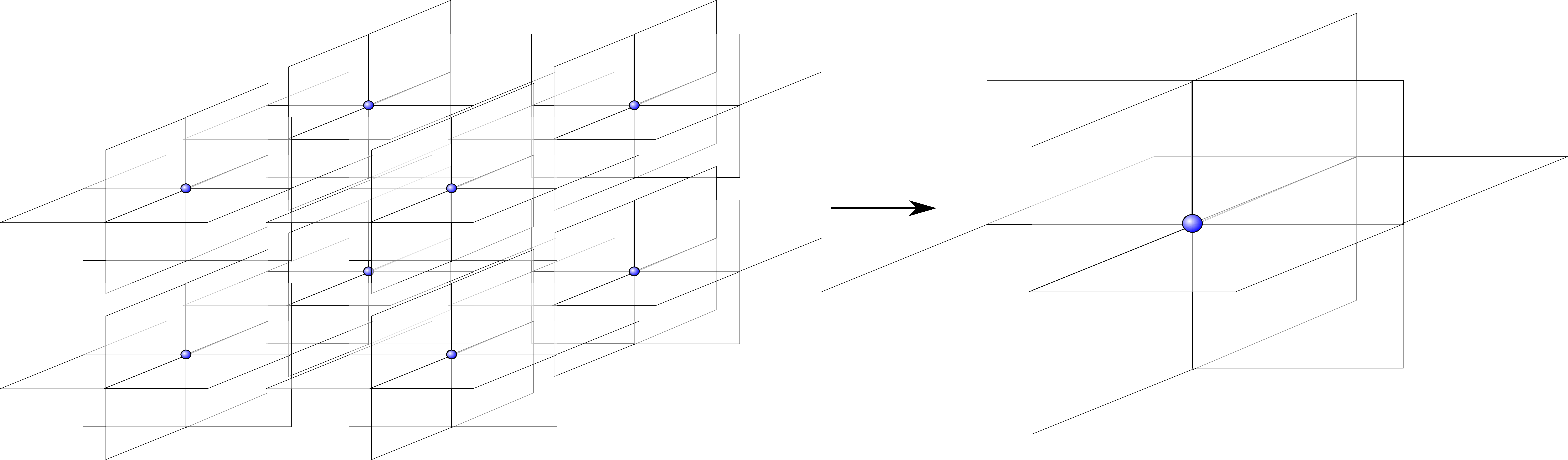}
	\caption{The RG flow rests on a coarse graining step, replacing several fine vertices by an effective coarse one.}\label{Fig:BIR_CoarseGrainCubicVertex}
\end{figure*}

To relate the amplitudes on the original vertex, and the new effective one, one needs an identification of degrees of freedom. This can be realized by a projection of configuration spaces, or injection of boundary Hilbert spaces
\begin{eqnarray}\label{Eq:EmbeddingMapsHilbertSpaces}
	\iota_{\Gamma'\Gamma}\;:\;\mathcal{H}_{\Gamma}\;\longrightarrow\;\mathcal{H}_{\Gamma'}.
\end{eqnarray}

\noindent The state sum (\ref{Eq:GeneralSetup_StateSum}) on the fine lattice with the fine boundary $\Gamma'$ has a dynamics which is given by a fine amplitude $\mathcal{A}_{\Gamma'}$. We make the ansatz for the fine amplitude to be of the EPRL type, i.e.~a local expression over the vertices (\ref{Eq:GeneralSetup_StateSum}), which gives a fine amplitude $\mathcal{A}_{\Gamma'}\;:\;\mathcal{H}_{\Gamma'}\;\to\;\mathbb{C}$, by
\begin{eqnarray}\label{Eq:BIR_RefineAmplitude}
	\mathcal{A}_{\Gamma'}(\psi_{\Gamma',j_f,\iota_e})\;:=\;\sum_{j'_f,\iota'_e}\prod_v\widehat{A}_{v},
\end{eqnarray}

\noindent where the sum ranges over all bulk spins and intertwiners $j'_f$, $\iota'_e$ of the fine lattice, while the boundary spins and intertwiners $j_f$,$\iota_e$ are kept fixed. The amplitude $\mathcal{A}_{\Gamma'}$ contains all information of the fine theory, represented as amplitude on the fine boundary. The renormalized amplitude is then given by
\begin{eqnarray}\label{Eq:BIR_RenormAmplitude}
	\mathcal{A}_{\Gamma}^{(\text{ren})}\;=\;\mathcal{A}_{\Gamma'}\;\iota_{\Gamma'\Gamma}.
\end{eqnarray}

\noindent In the case of a nested hypercubic lattice, note that the vertex amplitudes in (\ref{Eq:BIR_RefineAmplitude}), as well as the renormalized amplitude (\ref{Eq:BIR_RenormAmplitude}) can be regarded as amplitudes on the same boundary graph $\Gamma$. However, they do not necessarily have to coincide. This is the essence of the concept of ``running coupling'' in the RG flow. 

In fact, equation (\ref{Eq:BIR_RenormAmplitude}) defines the RG flow of the model, i.e.~the relation of amplitudes on different discretization scales. Mathematically, this is the notion of cylindrical consistency, which is required to define the continuum limit.\footnote{ This should not be confused with the notion of cylindrical consistency employed in the construction of the Ashtekar-Lewandowksi vacuum in loop quantum gravity, which is entirely kinematical (see e.g.~\cite{thomasbook}, and the discussions in \cite{Bahr:2011aa, Dittrich:2012jq, Bahr:2014qza, Dittrich:2014ala}).} Notably, assume one has solved the RG flow equations along all lattices, i.e.~one has a collection of amplitudes $\{\mathcal{A}_\Gamma\}_\Gamma$ which satisfy cylindrical consistency:
\begin{eqnarray}\label{Eq:BIR_CylindricalConsistency}
	\mathcal{A}_{\Gamma}\;=\;\mathcal{A}_{\Gamma'}\;\iota_{\Gamma'\Gamma}.
\end{eqnarray}

\noindent for all $\Gamma\leq\Gamma'$, i.e.~whenever $\Gamma$ arises as a refinement of $\Gamma'$. Then, this is a necessary condition that the continuum amplitude $\mathcal{A}_\infty:\mathcal{H}_\infty\to\mathbb{C}$ can be defined on the continuum Hilbert space
\begin{equation}
	\mathcal{H}_\infty\;:=\;\lim_{\Gamma\rightarrow \infty}\mathcal{H}_\Gamma,
\end{equation}

\noindent which is the inductive limit of all the $\mathcal{H}_\Gamma$. See \cite{Dittrich:2012jq} for details. 

This shows a nice interplay between mathematical concepts and physical intuition. The notion of scale is here played by the choices of lattices, and their relation to one another, which provide a hierarchy among the degrees of freedom. Note that, even though in our case the lattices are regular hypercubic ones, there are no lengths or other geometric properties assigned to them. Rather, the sum (\ref{Eq:BIR_RefineAmplitude}) ranges over different geometries \emph{of the same lattice}.

\subsubsection{On embedding maps}
It should be noted that the prescription depends on the way in which degrees of freedom are represented, and identified along different scales. In particular, the embedding map $\iota_{\Gamma'\Gamma}$ depends on these choices, which are not unique. For instance, any family of unitary operators $U_\Gamma$ on $\mathcal{H}_\Gamma$ lead to an equivalent theory with
\begin{eqnarray*}
	\tilde{\mathcal{A}}_\Gamma\;&:=&\;\mathcal{A}_\Gamma U_\Gamma,\\[5pt]
	\tilde{\iota}_{\Gamma'\Gamma}\;&:=&\;U_{\Gamma'}^{-1}\iota_{\Gamma'\Gamma}U_\Gamma.
\end{eqnarray*}

\noindent The precise choice of $\iota_{\Gamma'\Gamma}$ can make the actual problem of solving (\ref{Eq:BIR_CylindricalConsistency}) harder or easier. In particular, there are, in general, some choices which can work well -- or not so well -- in conjunction with certain approximation methods. 

In \cite{Dittrich:2012jq}, it is argued that the most beneficial way would be to use dynamical embedding maps, which in and of themselves already contain all the information of the dynamics of the theory. The reason for this is that one can interpret the embedding maps $\iota_{\Gamma'\Gamma}$ as ways to identify and add degrees of freedom under refinement. Then \eqref{Eq:BIR_CylindricalConsistency} suggests that refining should be done with respect to the dynamics encoded in the amplitude $\mathcal{A}_\Gamma$, i.e.~degrees of freedom should be added in the dynamical vacuum state. This is a highly non-trivial condition on both $\mathcal{A}_\Gamma$ and $\iota_{\Gamma'\Gamma}$. A real-space coarse graining algorithm, called tensor network renormalization \cite{levin,guwen,vidal-evenbly}, aims exactly at implementing such a scheme: the partition function of the system is rewritten as the contraction of a (local) network of tensors, which does not refer to a background and does not require a notion of scale. This network is coarse grained by defining effective coarse degrees of freedom from fine ones and ordering them by dynamical relevance. Thus these variable transformations, given by the dynamics, are the inverse of embedding maps. To keep this algorithm numerically feasible, one usually has to truncate the maximum number of degrees of freedom kept in each iteration. In quantum gravity, this algorithm has been successfully applied to 2D analogue spin foam models for finite \cite{Dittrich:2011zh,Dittrich:2013bza} and quantum groups \cite{Dittrich:2013voa,Steinhaus:2015kxa,Dittrich:2016tys} and 3D lattice gauge theories \cite{Dittrich:2014mxa,Dittrich:2016tys}. One of its main advantages is the applicability to oscillating amplitudes and fermionic systems \cite{Banuls:2016jws}. However a main disadvantage is the exponential growth in numerical cost with growing number of degrees of freedom, which has prohibited a direct application to 4D spin foam models.

When using the physical embedding maps, the continuum Hilbert space is equivalent to the physical Hilbert space, in which time translation becomes trivial, i.e.~scattering matrix elements are simply computed taking the inner product between in- and out-states.

Since we do not have the physical embedding maps at our disposal (indeed they would have to be found by solving the RG flow equations), we instead use an ad hoc choice for embedding maps, which identify (kinematical) geometric quantities among different scales, such as spins.  The degrees of freedom here are added by $\iota_{\Gamma'\Gamma}$ in such a way that e.g.~fine areas add up to coarse areas. This condition is translated to a condition on the coupling of fine spins to coarse spins. Details can be found in \cite{Bahr:2017klw}.

\subsubsection{Projected RG flow}

In general, the cylindrical consistency equations (\ref{Eq:BIR_CylindricalConsistency}) are very hard to solve, even though we have restricted ourselves to specific lattices\footnote{See \cite{Bahr:2014qza} for the treatment of a case allowing for all possible lattices at the same time. In that case, there are uncountably many RG flow equations to solve.}. To simplify matters, one can instead consider amplitudes $\mathcal{A}_{\Gamma}^{(g_i)}$ on $\Gamma$, which are given in terms of few parameters $g_i$, called coupling constants. One then attempts to rewrite the flow of amplitudes in terms of a flow of coupling constants
\begin{eqnarray}
	g_i\;\longrightarrow\;g_i' \; .
\end{eqnarray}

\noindent The question whether a parametrization in terms of few coupling constants is feasible, depends on its renormalizability, i.e.~on whether the effect of the integrated out degrees of freedom in (\ref{Eq:BIR_RenormAmplitude}) can be absorbed by a shift in the $g_i$. Whether quantum gravity is renormalizable or not, is still an open question. While it is often argued that the perturbative formulation is not \cite{Goroff:1985sz}, there are hints that there might exist a non-Gaussian fixed point, around which the flow might be renormalizable \cite{Niedermaier:2006wt}. 

We have to leave this question open for now. To be able make computations, however, we truncate the flow to only finitely many parameters. That is, we make an ansatz for $\mathcal{A}_\Gamma=\mathcal{A}_\Gamma^{(g_i)}$ in terms of the EPRL model (\ref{Eq:GeneralSetup_EPRL}), with free parameters
\begin{eqnarray}\label{Eq:BIR_CouplingConstants}
	\{g_i\}\;=\;\{\alpha,\,G,\,\Lambda\},
\end{eqnarray}

\noindent i.e.~the parameter defined in the face amplitude (\ref{Eq:GeneralSetup_FaceAmplitude}), as well as Newton's constant $G$ and the cosmological constant $\Lambda$. We specifically do not choose the Barbero-Immirzi-parameter $\gamma$ as a running coupling, since its connection to the allowed spins is rather pathological in the Euclidean EPRL model. The precise range of allowed spins $k_f$ sensitively depends on $\gamma$, by the condition that $j_f^\pm$ given by (\ref{Eq:SpinsSimplicity}) are half-integers. In particular, changing $\gamma$ by a tiny amount can make huge changes in the range. In particular, the chosen boundary data which works for one $\gamma$ might not be allowed for another, which would spoil the RG flow equations. To avoid this complication, we fix the value to
\begin{eqnarray}
	\gamma\;=\;\frac{1}{2}.
\end{eqnarray}

\noindent Since the same pathology does not appear in the Lorentzian signature model, we surmise that, in that case, it would be prudent to also choose $\gamma$ as a running coupling. 

The projection of the flow will be achieved the following way: in its general form, the RG flow equation (\ref{Eq:BIR_CylindricalConsistency}) can be rephrased as the fact that all observables $\mathcal{O}_\Gamma$, which are defined on the coarse lattice $\Gamma$, can also be measured on the fine lattice $\Gamma'$ (where we denote them as $\mathcal{O}_{\Gamma'}=(\iota_{\Gamma'\Gamma})_*\mathcal{O}_\Gamma$), and the expectation values, respectively obtained with the amplitude $\mathcal{A}_\Gamma$ and $\mathcal{A}_{\Gamma'}$, agree, i.e.
\begin{eqnarray}\label{Eq:BIR_CylindricalConsistencyObservables}
	\langle \mathcal{O}_\Gamma\rangle_\Gamma\;=\;\langle\mathcal{O}_{\Gamma'}\rangle_{\Gamma'}\qquad\text{for all }\mathcal{O}_\Gamma.
\end{eqnarray}

\noindent  We emphasize that this is an equivalent rewriting of (\ref{Eq:BIR_CylindricalConsistency}). If we truncate the theory space to amplitudes given in terms of few coupling constants $g_i$, we cannot expect (\ref{Eq:BIR_CylindricalConsistencyObservables}) to hold for all observables any more exactly. Instead, we will only demand it to hold approximately, for a subset of all observables. In particular, we choose a finite set of observables $O_\Gamma^{(n)}$, which we call \emph{reference observables}, and demand that the error
\begin{eqnarray}
	\Delta_{\Gamma,\Gamma'}^{g,g'}\;:=\;\sum_{n}\,\big| \langle \mathcal{O}_\Gamma^{(n)}\rangle_\Gamma^{g}\,-\,\langle\mathcal{O}_{\Gamma'}^{(n)}\rangle_{\Gamma'}^{g'}   \big|^2
\end{eqnarray}

\noindent is minimal\footnote{To simplify notation we refer to all parameters by $g=(\alpha,G,\Lambda)$ and drop the subscript $i$.}. This truncation of the RG flow obviously depends in the choice of observables, and a good flow requires that one finds observables which capture the dynamics of enough interesting degrees of freedom. 

In this article, we choose a specific set of observables, depending on the situation we are in. We will describe these in more detail in section \ref{Sec:IsotemporalRGFlow}. In particular, we will, in some instances, truncate the flow further and keep some of the parameters in (\ref{Eq:BIR_CouplingConstants}) fixed. Depending on which and how many, the choice for reference observables will be adapted.

\subsection{Approximations}\label{Sec:Approximations}
In order to solve the RG flow equation \eqref{Eq:BIR_CylindricalConsistencyObservables} we adopt a number of approximations 
\begin{itemize}
	\item {\itshape Reduced state sum}: The partition function \eqref{Eq:GeneralSetup_StateSum}  is hardly usable to carry out predictions about transition probabilities and expectation values of observables. This fact has roots in the complexity of its expression which involves a sum over all the possible geometric configurations $\{j_f,\iota_e\}$. To overcome this issue we restrict the state sum to a special set of symmetric configurations. They define a discretization of spacetime in which only a limited number of spins $j_f$ is required to keep track of the geometric degrees of freedom, while all the intertwiners $\iota_e$ are confined into the shape of a so-called {\itshape quantum frustum}. We will describe this structure in more details in the next section. 
	\item {\itshape Semiclassical limit}: In the large spin limit the EPRL-FK vertex amplitude \eqref{Eq:VertexTrace} has been proven to be connected to discrete GR, when built on a simplicial discretization \cite{Barrett:2009gg}. This result was confirmed in \cite{Bahr:2017eyi} by a saddle point approximation of the reduced amplitude. As we will see, unlike the case of a general simplicial decomposition, the thinning of the state sum leads to an explicit asymptotic expression of \eqref{Eq:RepackagedSpinFoamSum} as a function of the spins. 
	This allows us to numerically evaluate the expectation values \eqref{expvalG} for some geometric observables  $\mathcal{O}_{\Gamma}$ on a given boundary graph $\Gamma$. Although the error one makes by replacing the amplitude with its large-$j$-asymptotic expression is hard to estimate, it can be expected that the approximation is quite good already for small values of the spins \cite{Bayle:2016doe, Bahr:2017eyi}. Since for large parts of the phase space the multi-vertex-amplitude appears to be suppressed for small spins \cite{Bahr:2015gxa}, the error might in fact not be that large. Still, this point warrants further study.
	\item {\itshape Projection of the amplitudes}: 
	In general, given a theory defined by a set of couplings $g_i$, the dimension of the parameter space can grow or decrease when one looks at the physics at different scales. In other words, new parameters may arise during the coarse graining process. Here we truncate the RG flow by considering the system as self-similar at all the scales. Thus, at each renormalization step we project the amplitude down to the reduced Euclidean EPRL-FK model defined by three parameters $g_i=(\alpha,G,\Lambda)$.
	 Again, we remind the reader that due to this, equation  (\ref{Eq:BIR_CylindricalConsistency}) can at best be satisfied approximately, see the previous section.
\end{itemize}

\begin{figure}[h]
	\centering
	\includegraphics[scale=0.6]{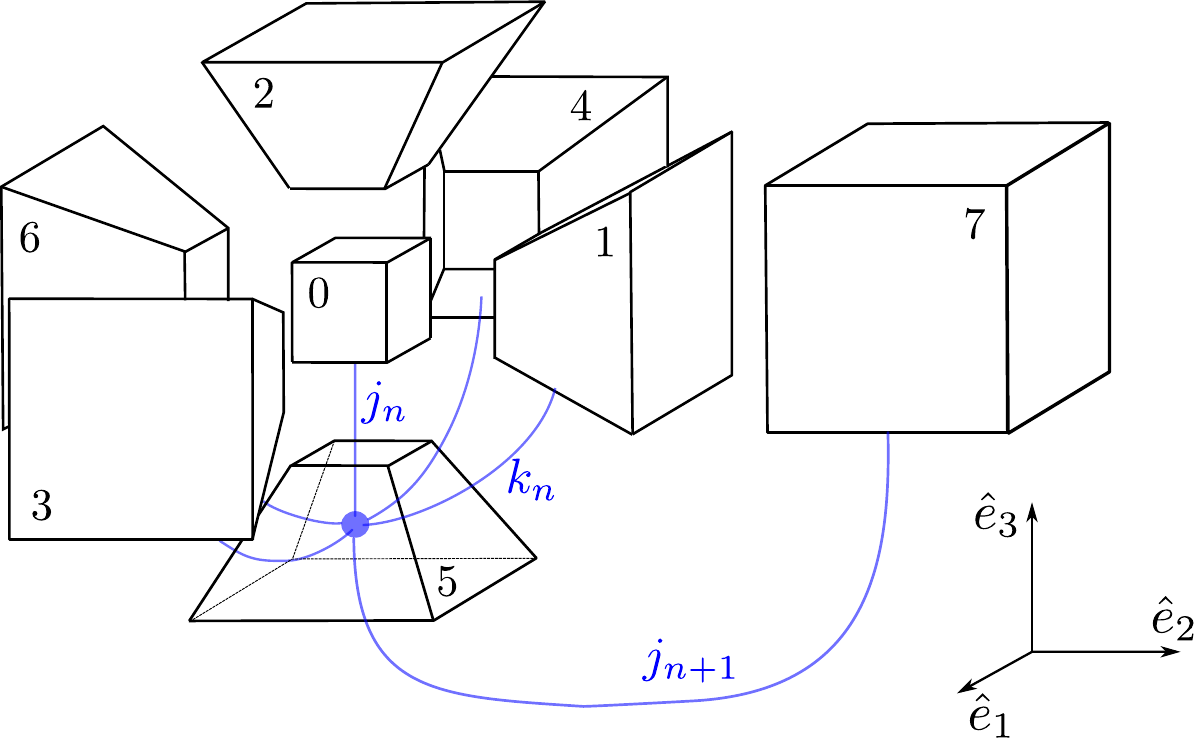}
	\caption{
		\small
		In black, the 3d boundary of an hyperfrustum $\mathrm{F}_n$. In general, we can think of it as the one-time-step evolution of a 3d boundary cube. In blue, the six-valent node dual to a boundary frustum $\mathrm{f}_n$. A similar node is associated to each hexahedron in figure.}
	\label{fig:boundhyper}
\end{figure}

The above set of approximations has been proven successful in some recent papers \cite{Bahr:2016hwc,Bahr:2017klw} where the use of a discretization in terms of hypercuboids allowed the evaluation of the RG flow of the parameter $\alpha$ appearing in the face amplitude. Also, the detection of a UV-attractive fixed point $\alpha_c$ showed an indication of invariance of the model under refinement. While opening the way to the numerical study of the continuum limit of restricted spin foams, the hypercuboid model stands on a severe restriction of d.o.f. which does not allow for curvature. The curvature is in fact vanishing everywhere and thus the theory is independent of other interesting parameters such as Newton's constant $G$ and the cosmological constant $\Lambda$. 
In this article we will instead work on a discrete structure introduced in \cite{Bahr:2017eyi} and specially designed to support a basic concept of curvature. 
It consists of a {\itshape pyramidal} discretization that, in the limit of large refinement, provides a natural description of a foliated manifold $\mathcal{M}=\Sigma\times \mathbb{R}$ in which the spatial hypersurfaces $\Sigma\sim T^3$ have the topology of a 3-torus, are flat, isotropic and homogeneous and can grow or contract at successive times. 
The typical grain of spacetime, defining the spin foam vertex, is the so called {\itshape hyperfrustum} $\mathrm{F}_n$ i.e., the four dimensional generalization of a truncated regular square pyramid (to which we refer as {\itshape frustum}).
We represent it in Fig.\ref{fig:boundhyper} via its 3d boundary, obtained by unfolding $\mathrm{F}_n$  into six equal frusta $\mathrm{f}_n$  and two cubes $\mathrm{c}_n$ and $\mathrm{c}_{n+1}$ of different sizes \footnote{This is the analogue, one dimension higher, of the unfolding of a 3d frustum into four regular trapezoids and two squares.}.
The geometry of the hyperfrustum is fully specified by three spins i.e.,  $\mathrm{F}_n= \mathrm{F}_n(j_n, j_{n+1}, k_n)$. 
The spatial spin $j_n$, corresponding to the face areas of $c_n$, determines the scale factor $a_n=\sqrt{G j_n}$ at a fixed time $t_n$, where $G$ is Newton's constant. The height of the hyperfrustum, defined as the distance between the centres of its boundary cubes, determines instead the time step $H_n=H_n(j_n,j_{n+1},k_n)=t_{n+1}-t_n$, where $k_n$ are the time-like spins of the trapezoidal faces of $\mathrm{f}_n$.

The reader familiar with spin foam models might be puzzled by our setup, where we claim to allow for discrete geometries with curvature while using spin foam amplitudes in the large-$j$-limit. Indeed, this limit is the context in which the so-called ``flatness-problem'' was discovered and discussed in great detail \cite{Bonzom:2009hw,Hellmann:2013gva}. It states that in this limit, no matter the boundary state of the spin foam, the bulk geometry is flat and accidental curvature constraints occur. In our case, where we only study a subset of the full spin foam path integral, the configurations that we permit in principle allow for curvature, in particular compared to the previously studied cuboid configurations. From our numerical studies, which we report in this article, we do not observe that this subset of the path integral is dominated by flat, i.e.~cuboid, geometries. Due to the restrictiveness of the path integral studied here, this finding is by no means a proving that the flatness problem is non-existent, yet it hints towards its intricacies that we need to understand better.

As it is shown in \cite{Bahr:2017eyi}, the model characterizes a cosmological subsector of the quantum theory in the sense that the associated {\itshape classical} Regge action reproduces the dynamics of a FLRW Universe on large refined discretizations.  In a {\itshape quantum} regime, the results of this model can potentially approximate the properties of a region of the Universe in which the dominating quantum fluctuations manifest the same symmetries of the Friedmann cosmology. Whether such systems may exist or not is not clear. Nonetheless, the interest in the model lies beyond the application to cosmology, since one can use it to investigate its RG flow. 

\begin{figure}[h]
	\centering
	\includegraphics[scale=0.38]{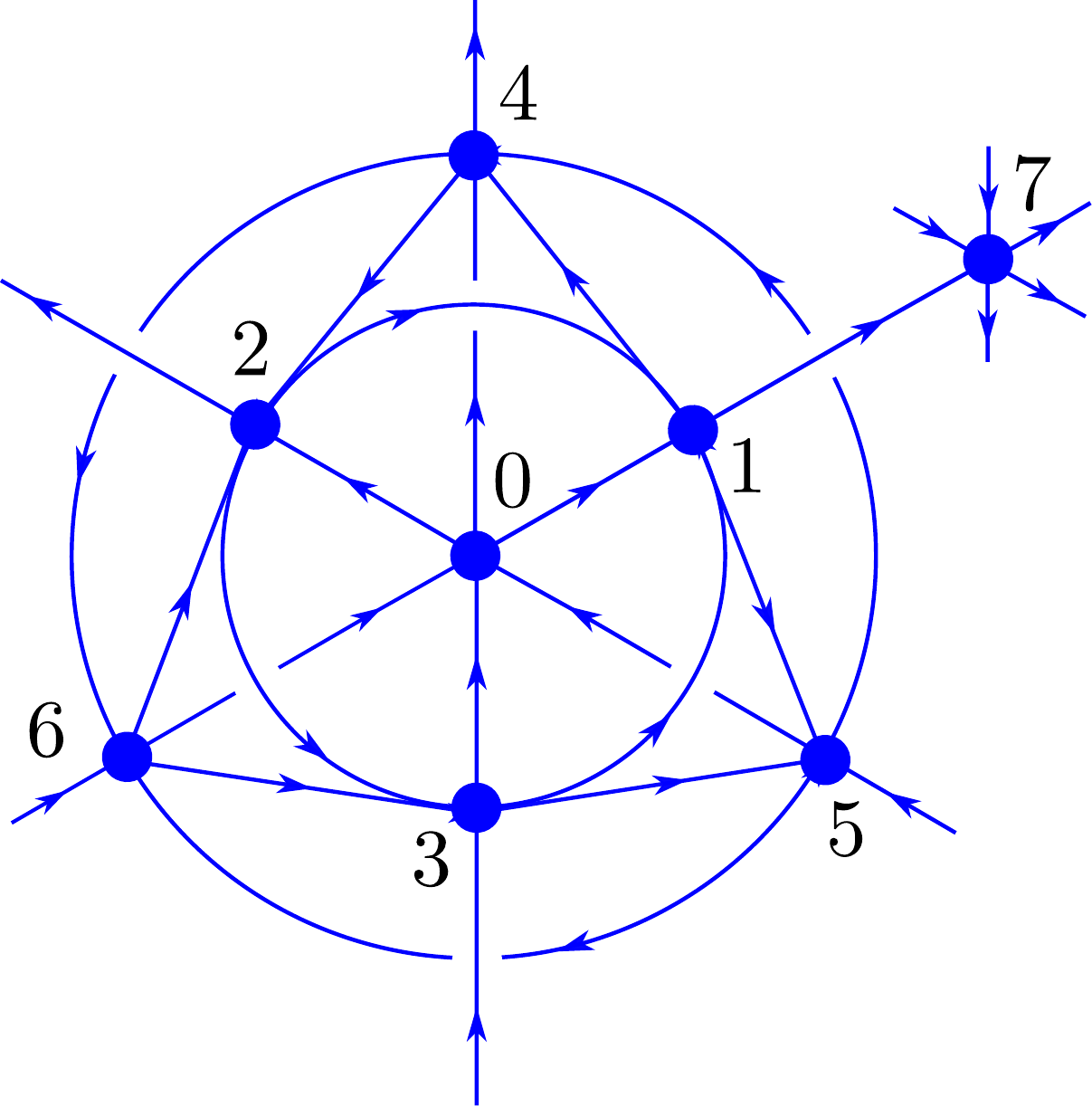}
	\caption{Spin network graph dual to a hyperfrustum boundary. The external legs are considered to be connected to the node $a=7$.
		\small
	}
	\label{fig:snf}
\end{figure}

The data to build the reduced EPRL-FK vertex amplitude is stored in the spin network dual to the boundary of a hyperfrustum (Fig.\ref{fig:snf}). This consists of eight six-valent nodes  $a=0,1,\cdots,7$ laced through their links $ab$. An intertwiner $\iota_a$ is assigned to each node $a$ and a spin $j_{ab}$ is attached to each link connecting the nodes $a$ and $b$. This labeling endows the (so far just combinatorial) graph with a geometric connotation so that, whenever two nodes share a link, two boundary hexahedra have the same face area bound.

The intertwiner at a node generically describes a {\itshape quantum frustum}. In the notation of Fig.\ref{fig:boundhyper} we can write for example the intertwiner at the node $a=5$ as  
\begin{equation}\label{coherentint}
	\iota_5=\int \mathrm{d}  g \ g   \ \triangleright \Bigg(| j_n , \hat{e}_3\rangle \otimes  | j_{n+1} , - \hat{e}_3\rangle 
	\otimes \bigotimes_{l=0}^3 | k_n , \hat{r}_l\rangle \Bigg),
\end{equation}
where $\hat{r}_l \equiv e^{- i \frac{\pi}{4} l \sigma_3}e^{- i \frac{\phi}{2} \sigma_2}\triangleright\hat{e}_3$ ($l=0,1,2,3$) are the four vectors perpendicular to the side faces of the frustum, while the slope angle $\phi$ between the top and the side face is a function of the spins
\begin{equation}\label{anglespin}
	\cos \phi=\frac{j_{n+1}-j_n}{4k_n} .
\end{equation}
This object reduces to a quantum cuboid for $j_{n+1}=j_n$.
In terms of these coherent states the vertex amplitude \eqref{Eq:VertexTrace} for $\gamma<1$ factorizes as $\mathcal{A}_{v}=\mathcal{A}_{v}^{+} \mathcal{A}_{v}^{-}$ being
\begin{equation}\label{vampl}
	\mathcal{A}_{v}^{\pm}=\int_{SU(2)^{8} } \mathrm{d} g_{a}^{\pm} e^{S_{\pm}[g_{a}^{\pm}]} ,
\end{equation}
the exponential of the complex action
\begin{equation}\label{act}
	S_{\pm} [g_{a}^{\pm}]= \frac{1}{G}\frac{|1 \pm \gamma |}{2} \sum_{ab \supset a} 2 A_{ab} \ln \langle- \vec{n}_{ab}|(g_{a}^{\pm})^{-1}g_b^{\pm}|\vec{n}_{ab}\rangle,
\end{equation}
where we denote the area $A_{ab}=Gj_{ab}$\footnote{We work in units in which $\hbar=1$, so $G=\ell^2_{\text{Planck}}$ has the dimension of an area. }, and $|\vec{n}_{ab}\rangle \equiv |1/2,\vec{n}_{ab}\rangle,$ and we call $\vec{n}_{ab}\in S^2 \subset \mathbb{R}^3$ the vector orthogonal to the face of $a$ which is dual to the link $ab$ (see Fig.\ref{fig:boundhyper} for a reference).
In the large spin limit the amplitude \eqref{vampl} behaves as an highly oscillatory integral and we can evaluate it via stationary phase approximation. 
Thus, we only consider the $SU(2)$ group elements $g_a$ that contribute most in the asymptotic limit i.e., the stationary and critical points such that $\partial S|_{g_a}=0$ and $\mathrm{Re} S(g_a) = 0$. The critical point equation can then be written as
\begin{equation}\label{criteq}
	g_a^{\pm} \triangleright \vec{n}_{ab}=- g_b^{\pm}\triangleright \vec{n}_{ba}.
\end{equation}
There are four distinct solutions for the couple $(g_a^{+},g_a^{-})$, corresponding to the set of rotations such that the boundary hexahedra are glued together at their faces so to reconstruct a 4d hyperfrustum.
The same method can be used to compute the norm of the coherent intertwiners to the leading order in the large j limit, and thus the edge amplitudes. 

These calculations were performed in a preparatory article \cite{Bahr:2017eyi} where the asymptotic limit of the full dressed vertex amplitude is carried out explicitly. 

Furthermore, in this article we also implement the cosmological constant $\Lambda$ by using a generalization of Han's deformation of the EPRL-FK model \cite{Han:2011aa}. This model is defined in \cite{Bahr:2018ewi}. The quantum amplitude is deformed not (as is usual) by replacing $SU(2)$ by its quantum counterparts, but rather by introducing specific operators for crossings in the boundary graph, which rely on a quantization of the classical formula for the $4$d-volume of a polyhedron \cite{Bahr:2017ajs}.
\footnote{The deformation of the vertex amplitude depends on an additional parameter $\omega$, see again \cite{Bahr:2018ewi}. In the asymptotic formula this parameter appears in the action in front of the dimensionless volume term, i.e.~the volume expressed in terms of spins. Thus $\omega$ is dimensionless and related to the cosmological constant via $\omega \sim \Lambda G$. Expressing the action in terms of areas results in eq. \eqref{vamplj}.}. 

Although nontrivial, it can be shown that, remarkably, the large-$j$-asymptotics of this deformed amplitude is rather similar to the undeformed case, in that both the position of the critical and stationary points, as well as the Hessian matrix are unchanged by this deformation process. Hence, the asymptotic action is simply  amended by a cosmological constant term from Regge Calculus \cite{Bahr:2018ewi}
\footnote{This is one example of how to incorporate a cosmological constant in SFM. Other methods were developed before \cite{Smolin:1995vq,Major:1995yz,Borissov:1995cn,Turaev:1992hq,Noui:2002ag,Han:2010pz,Fairbairn:2010cp,Haggard:2014xoa,Dupuis:2013haa}, e.g.~by replacing the Lie group by a quantum group.}.

In order to explicitly write the amplitudes in a compact form let us first define the functions
\begin{equation}\label{short}
	\begin{split}
		\Omega&\equiv \frac{\sqrt{1-\gamma^2}}{8\pi}, \qquad \qquad Q\equiv 2+\frac{j_n+j_{n+1}}{2k_n},\\
		\theta&\equiv\arccos \frac{1}{\tan\phi}, \qquad \ \ K\equiv \sqrt{-\cos 2\theta},
	\end{split}
\end{equation}
where the slope angle $\phi$ is given in \eqref{anglespin}.
Then, for a face $f$ labeled by the spin $j_n$ (similar for $k_n$), for an edge $e$ dual to a frustum $\mathrm{f}_n(j_n,j_{n+1},k_n)$ and for a vertex $v$ dual to a hyperfrustum $\mathrm{F}_n(j_n,j_{n+1},k_n)$ the asymptotic expressions of the respective amplitudes are 
\begin{equation}\label{fam}
	\mathcal{A}_{f} \rightarrow (8\pi\Omega j_n)^{2\alpha},
\end{equation}
\begin{equation}\label{eamplf}
	\mathcal{A}_{e}\rightarrow  \frac{\Omega^3 k_n^3}{2(4\pi)^4} (1+K^2)(1+K^2-2Q)^2,
\end{equation}
\begin{equation}\label{vamplj}
	\mathcal{A}_v \ \rightarrow \Big(\frac{1}{8\pi\Omega}\Big)^{21} \Big(\frac{e^{i \frac{S_{R}}{G} }}{-D}+\frac{e^{-i\frac{S_{R}}{G} }}{-D^{*}}+2\frac{\cos ( \frac{ \gamma S_{R}} {G}-\frac{\Lambda}{G}\mathcal{V})}{\sqrt{D D^{*}}}\Big).
\end{equation}
We recognize the Regge action $S_R=\sum_{h}A_h\epsilon_h$, which usually appears in the asymptotic limit of the spin foam model under consideration \cite{Barrett:2009mw, Conrady:2008mk, Bahr:2015gxa, Dona:2017dvf, Bahr:2017eyi}. It is a function of the areas $A=Gj$ via
\begin{equation}\label{reggea}
	S_R= G \left( 6j_n \Big(\frac{\pi}{2}-\Theta\Big) + 6j_{n+1} \Big(\frac{\pi}{2}-\Theta'\Big) +12k_n\Big(\frac{\pi}{2}-\Theta''\Big) \right).
\end{equation}
The four dimensional dihedral angles $\Theta_{ab}$ among the 3d blocks at the vertex boundary are (always refer to Fig.\ref{fig:boundhyper})
\begin{equation}
	\begin{split}
		\Theta=\theta \qquad &\mbox{if} \ \ \ a=0 \ \mbox{or} \ b=0 \; ,\\
		\Theta'=\pi-\theta \qquad &\mbox{if} \ \ \ a=7 \ \mbox{or} \ b=7 \; ,\\
		\Theta''=\arccos(\cos^2\theta) \qquad &\mbox{if} \ \ \  a,b \in\{1,\cdots,6\} \; .
	\end{split}
\end{equation}
Let us notice that in our symmetry restricted setting the use of spin variables is equivalent to the use of edge lengths in standard Regge calculus i.e., there is a unique invertible relation $j\leftrightarrow l$ which holds for any number of vertices glued together \cite{Bahr:2017eyi}. 
The cosmological constant term in the discrete setting is proportional to the four volume of the hyperfrustum
\begin{equation}\label{Vol}
	\mathcal{V}= G^2 \, k_n^2 \,  K \, (Q-2).
\end{equation}
The function $D=D(j_n,j_{n+1},k_n)$ is the determinant of the Hessian of \eqref{act} and its explicit expression is
\begin{equation*}\label{determH}
	\begin{split}
		D&=\frac{j_n^3j_{n+1}^3k_n^{15}}{16} K\left(K-i K^2+iQ\right)^3 \left(1+K^2-2 Q\right)^3 \\
		& (K+i)^6 (K-3 i)^2 \left(1+3 K^2-2Q-2 i K (Q-1)\right)^3.
	\end{split}
\end{equation*}
Finally, arranging the above function as $D=|D|\exp(i\varphi)$ and summing up all the contributions we can write the dressed vertex amplitude \eqref{Eq:RepackagedSpinFoamSum} as 
\begin{equation}\label{drevampl}
	\widehat{\mathcal{A}}_{v}\sim\frac{(j_n j_{n+1})^{3\alpha-\frac{3}{2}}k_n^{6(\alpha-1)}}{B} \Big(\cos( \frac{S_R}{G}+\varphi)+\cos(\frac{\gamma S_R}{G} -\frac{\Lambda}{G}\mathcal{V})\Big),
\end{equation}
with 
\begin{equation*}
	B=\frac{|D|}{\left(1+K^2\right)^3 \left(1+K^2-2 Q\right)^6}.
\end{equation*}
The first cosine appearing in the dressed vertex amplitude is sometimes addressed as `weird' being an unexpected term appearing in the asymptotics of the Euclidean EPRL-FK vertex amplitude \cite{Barrett:2009gg}. The argument of the second cosine is the correct Regge action with the proper cosmological constant term. The fact that it appears in a cosine  instead of an exponential is related to the so-called cosine problem. Despite the debate around the asymptotic limit of the EPRL-FK model, here we compute expectation values of observables with respect to this amplitude. This may shed a new light on the properties as well as the problems of the model. All the techniques used can be applied straightforwardly to other kind of amplitudes (e.g. without weird terms). 

In the next sections we use the fact that in the large spin limit the sum over the spins is well approximated by an integral $\sum_j \rightarrow \int \mathrm{d}j$ so that, given a boundary graph $\Gamma$, we can numerically integrate the observables $\mathcal{O}_{\Gamma}$ weighted with the dressed vertex amplitude \eqref{drevampl} and thus evaluate their expectation values
\begin{equation}\label{expvalG}
	\langle\mathcal{O}_{\Gamma}\rangle_{\Gamma}= \frac{\int \mathrm{d}j_{f} \mathcal{O}_{\Gamma} \prod_{v} \widehat{\mathcal{A}}_{v}}{\int \mathrm{d}j_{f} \prod_{v} \widehat{\mathcal{A}}_{v}}.
\end{equation}
Eventually, we use this to define and numerically solve the RG flow equation \eqref{Eq:BIR_CylindricalConsistencyObservables}.

One should note that physically one is actually integrating over areas $A$, rather than spins $j$, which are related by $A=Gj$, since we work in units in which $\hbar=1$. Still, in order to keep in line with the majority of the literature, we will, from now on, substitute $Gj \rightarrow j$, and also change the corresponding notations. Thus the spins have a physical dimension of areas and the explicit dependence on $G$ in \eqref{reggea} and \eqref{Vol} disappears. The overall integration measure acquires additional powers of $G$ as a factor, which does not play a role in the path integral nor in expectation values of observables. Eventually the state sum will depend on the three parameters $\alpha$, $G$ and $\Lambda$ as stated in \eqref{drevampl}.\footnote{Remember  that we have fixed $\gamma=\frac{1}{2}$. In general, the amplitude also depends on the Barbero-Immirzi parameter $\gamma$.}

\subsection*{Degrees of freedom}

It is worthwhile to recap which degrees of freedom we are summing over at this point.

Originally, the spin foam model depends on spins $j$ and intertwiners $\iota$. The truncation leaves us with a subset of variables $j_n$ and $k_n$ of spins (i.e.~areas), which are assigned to space-like and time-like faces in the $4d$ lattice, where the first ones describe the geometry of the isotropic and homogenous space-like Cauchy-surfaces, while the latter describe the transitions between hypersurfaces, i.e.~time-steps.\footnote{Remember though, that the choice of time-direction is somewhat arbitrary at this point, since we deal with Riemannian geometries in this article.} 

By going over to continuous areas, and because of the equivalence to length variables, this describes essentially a subsector of the state space of quantum Regge calculus. There are a few differences though: Firstly, the factor coming from the Hessian of the asymptotic formula induces a different measure. Secondly, the amplitude is not of the form $\exp(-S)$, but rather (\ref{drevampl}), i.e.~$\cos(\tilde{S})+\cos(S)$.

Also, it should be noted that the RG flow is defined slightly differently here, since we do not introduce a correlation length, but use, as ordering parameter, a different observable, usually certain volume fluctuations. These will be described in more detail in the following chapter. How to define a correlation length, other than in the pure combinatorial sense, is not obvious, but intriguing to explore in future research.

In principle, the integral over degrees of freedom is unbounded, which could lead to divergencies of the integral in the limit of large spins $j,k\to\infty$. However, depending on the value of the coupling constant $\alpha$, the integrand goes to zero sufficiently fast in that limit, so the integral stays finite. This has been discussed for hypercuboids in \cite{Bahr:2015gxa}, and a similar calculation is true for the frustum case, which we consider in this article. In particular, we only consider a flow of $\alpha$ well inside the region in which the large-$j$-region is not a problem.

The Hessian matrix which occurs in the measure factor of the path integral goes to zero in the limit of vanishing spins, which might a priori lead to divergencies in the $j,k\to 0$ region as well. However, this is an artefact of the asymptotic formula, which does not hold for the small spin case. Indeed, the actual amplitude stays finite in that region, where the integral would have to be replaced by the sum anyway. Indeed, our numerical investigations show that there is usually only a very small region around $j,k\approx 0$ in which the amplitude diverges. Figure (\ref{Fig:Iso_IntegrandAtFixedPoint}) is an example for this behaviour, in which we find that the integrand itself tends to zero as spins approach small values, and only suddenly diverges very close to $j,k= 0$. We attribute this behaviour to the breakdown of validity of the asymptotic formula, and remove it by introducing a small spin cutoff. As long as one does not enter the region in which the asymptotic formula breaks down anyway, the results appear not to be influenced by the precise position of the cutoff.

\subsection{Numerics}\label{Sec:Numerics}

The vital ingredient of this article is the calculation of expectation values of geometrical observables in the spin foam state sum. The spin foam amplitudes are intricate functions of the spins $j$, and the integrations over $j$ generically cannot be performed analytically. As in a similar analysis for cuboid-shaped spin foams \cite{Bahr:2015gxa, Bahr:2016hwc, Bahr:2017klw} we will therefore perform these integrations numerically.

We perform our numerical simulations in the programming language \url{Julia}\footnote{\url{https://julialang.org/}} and use algorithms suitable for higher-dimensional integration from the \url{Cuba} package \cite{Hahn:2004fe}\footnote{See \url{https://github.com/giordano/Cuba.jl} for the package and documentation how to use these algorithms in \url{Julia}.}.

While the \url{Cuba} package contains several algorithms, most of which employ Monte Carlo techniques, we use a {\it deterministic} algorithm called \url{Cuhre}. It roughly works as follows: Similar to Monte Carlo algorithms, the integrand is evaluated at several points. Given this data, \url{Cuhre} then attempts to approximate the integrand by a polynomial in the integration variables and estimates the error. If the error is larger than requested, the region with the largest error gets subdivided and the algorithm is iterated. Once this procedure has sufficiently converged, or the maximum number of iterations has been reached, the polynomials are used to deterministically evaluate the integral.

For our purposes this algorithm is particularly useful since it is more efficient for integrating oscillatory integrands than ordinary Monte Carlo techniques, at least if the dimensionality of the integrand is not too high\footnote{To approximate higher dimensional regions by polynomials requires considerably more sample points rendering the algorithm less efficient.}. Indeed, as frusta configurations allow for curvature, the vertex amplitude is a sum of several oscillating terms, which marks an important generalisation compared to the pure cuboid case. Fortunately hyperfrusta are prescribed by only three spins, compared to six of a hypercuboid. Together with the large amount of symmetry in these configurations, we can study discretisations containing many spin foam vertices, which only depend on a few spins. Indeed most of the integrations performed in this article are two--dimensional, which can be efficiently performed.

Another generalisation compared to the cuboid case is the necessity of introducing a cut-off on the spins. While in the cuboid case we implemented an embedding map fixing the total area of a coarse face, we a priori cannot enforce such a restriction onto the hyperfrusta. To efficiently perform the integrals, an upper cut-off on the spins is necessary. Usually one then has to carefully check that the result does not change under gradually increasing the cut-off. In our case this question is closely tied to the value of the parameter $\alpha$ as it determines whether large or small spins are preferred in the path integral. Generically if $\alpha$ is too large the result is cut-off dependent as the amplitudes diverge for growing spins. We have performed our simulations in a regime of $\alpha$ where the results converge for relatively small cut-off $j_{\text{max}} \sim 10$. Fortunately this is also the regime of interesting dynamics.

Thus the difficulty of the numerics stems less from the integrand itself but from the fact that we have to scan a 3-dimensional parameter space. To quickly generate the results we have used the local HPC at Perimeter Institute, e.g. to perform 1024 one- and two-dimensional integrations took roughly 12 hours on a single core. This can be further accelerated as the \url{Cuba} package in \url{Julia} can be straightforwardly vectorized and parallelized.

\section{Renormalization group flows }\label{Sec:IsotemporalRGFlow}
We work on a system $\Phi$ describing the time evolution of an isotropic and homogeneous universe. We consider different discretizations of this process in terms of hyperfrusta. We then demand cylindrical consistency among different discretizations, which defines the RG flow of the amplitudes.

\begin{figure}[h]
	\hspace*{-0.5cm} 
	\includegraphics[scale=0.47]{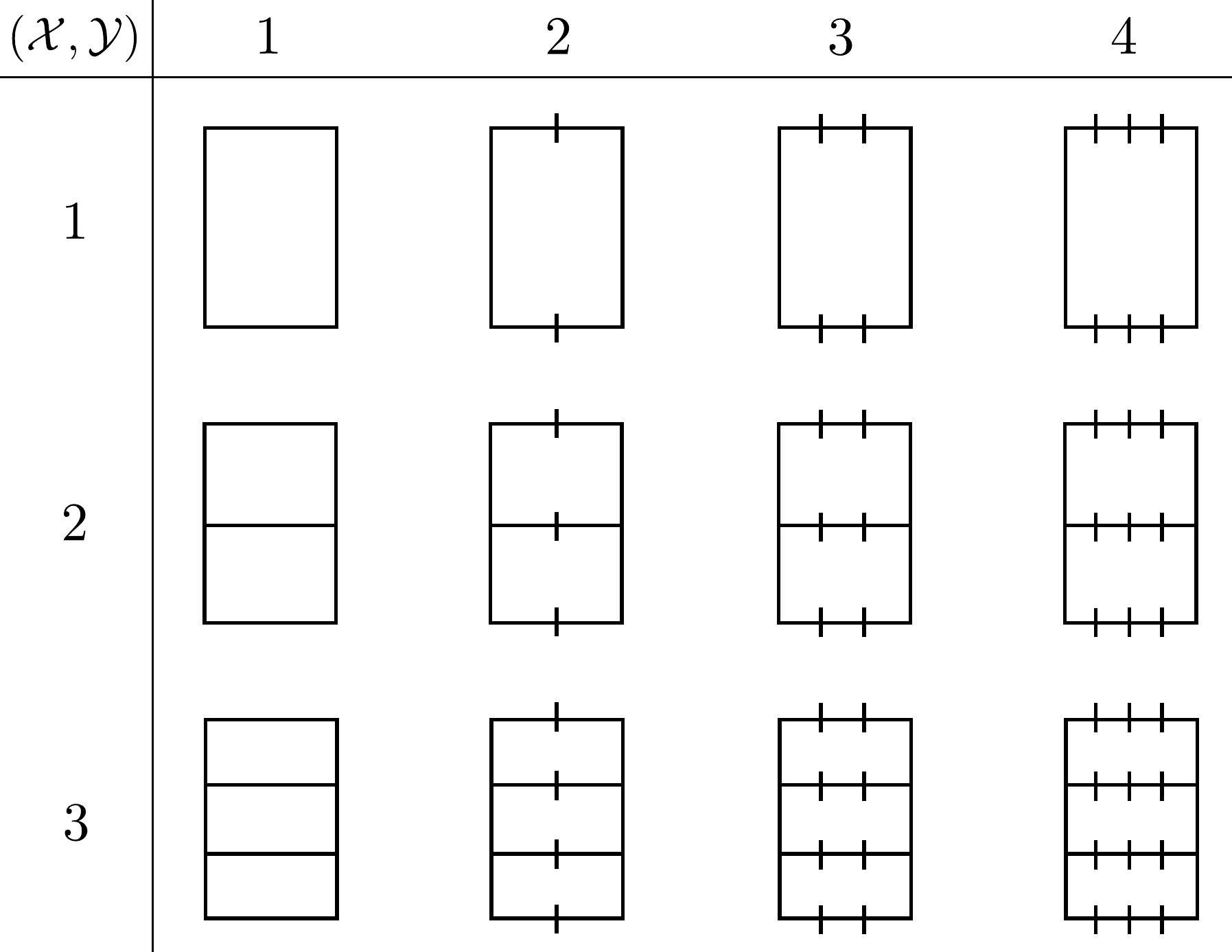}
	\caption{Catalog of some possible discretizations of $\Phi$ which preserve the homogeneity of the spatial hypersurfaces. The labels $\mathcal{X}$ refer to the number of links used to discretize each spatial direction. The labels $\mathcal{Y}$ refer to the number of time steps in which the transition occurs.}
	\label{fig:Discretizations}
\end{figure}

Let us focus on the table in Figure \ref{fig:Discretizations} which catalogs some possible discretizations of $\Phi$ preserving the symmetries of the system.  
Each slot $(\mathcal{X},\mathcal{Y})$ represents a discretization $\Phi_{(\mathcal{X},\mathcal{Y})}$ of $\Phi$ in terms of $n=\mathcal{X}^3\mathcal{Y}$ vertices.

In what follows we are considering the initial and final slices as our disconnected boundary.
There exists a unique embedding map \eqref{Eq:EmbeddingMapsHilbertSpaces} which allows for using only and solely the hyperfrustum vertex at each refinement step. This is such that it maps a coarse boundary cube into the unique configuration of $\mathcal{X}^3$ smaller cubes, all of the same size.


At the coarsest level the process is described by a single vertex i.e., a hyperfrustum $\Phi_{(1,1)}$ with boundary cubes of areas $j_i$ and $j_f$. These labels fix the boundary geometry of $\Phi_{(1,1)}$ and determine the coarsest scale where there is a single degree of freedom available e.g., the height $H$. 
Shifting to the right in the picture (i.e. along $\mathcal{X}$) corresponds to a homogeneous split of the spatial discretizations, dictated by the embedding map. 
Thus, in the slot $(\mathcal{X},1)$  each spatial edge is splitted into $\mathcal{X}$ equal pieces. Correspondingly, each of the coarsest boundary cubes of areas $j_i$ and $j_f$ is respectively subdivided into $\mathcal{X}^3$ cubes of areas $j_i/{\mathcal{X}^2}$ and $j_f/{\mathcal{X}^2}$. 
Stepping down in the picture (i.e. along $\mathcal{Y}$) corresponds instead to refining the discretization in the time direction. As an example, at the slot $(1,\mathcal{Y})$ of Table \ref{fig:Discretizations} one has the transition of a single cube in $\mathcal{Y}$ time steps which is represented by a chain of $\mathcal{Y}$ hyperfrusta of heights $H_1,\dots,H_{\mathcal{Y}}$ with $\sum_{i=1}^{\mathcal{Y}}H_i=H$.
The variables of a discretization $\Phi_{(\mathcal{X},\mathcal{Y})}$ are the bulk spatial spins $j_n$ and the time-like spins $k_m$, where $n=1,\ldots, \mathcal{Y}-1$, $m=1,\ldots \mathcal{Y}$.  

The flow is extrapolated from the comparison of the dynamics of two discretizations $\Gamma=\Phi_{(\mathcal{X},\mathcal{Y})}$ (coarse) and $\Gamma'=\Phi_{(\mathcal{X}',\mathcal{Y}')}$ (fine) defining a coarse graining step. One can choose whatever couple $(\Gamma,\Gamma')$ in Table \ref{fig:Discretizations} with the condition that $\mathcal{X}\cdot\mathcal{Y}<\mathcal{X}'\cdot\mathcal{Y}' $. In general, the flow will depend on such a choice. However we expect that for highly discretized $\Gamma$ and $\Gamma'$ the dependence of the flow becomes negligible since the discretization is fine enough to capture the dynamics of the system. 

Note that all the configurations shown in table \ref{fig:Discretizations} give rise to real transition amplitudes, since the vertex amplitudes $\widehat{\mathcal{A}}_{v}$  \eqref{drevampl} are real.
The discretizations laying in the even columns have positive amplitudes while the odd columns can take negative values since each time step comes with an odd power of vertex amplitudes.
In what follows we restrict ourselves to discretizations with positive amplitudes only. This ensures in general a faster numerical evaluation of the expectation values of the observables.


\subsection{One dimensional isochoric RG flow}\label{Sec:1DFlow} 
First, we consider a restricted flow where all coupling constants are kept fixed, except for $\alpha$. The RG flow in $\alpha$ is computed in the isochoric setting i.e., keeping fixed the total $4$-volume of space-time. This is a generalization of a previous work, in which the discretization has been restricted to hypercuboids, and where it has been observed that the $RG$ flow of $\alpha$ is intimately connected to the vertex displacement symmetry of the model \cite{Bahr:2016hwc}. 

In particular, in \cite{Bahr:2015gxa}, it was observed that the EPRL model breaks vertex displacement symmetry, which is the manifestation of diffeomorphisms on the lattice \cite{Bahr:2009ku, Bahr:2009qc,  Dittrich:2012qb,Dittrich:2014rha,  Bahr:2015gxa, Asante:2018wqy}. While this breaking of symmetry is well-known in classical Regge calculus, where it appears whenever curvature is involved, the quantum theory breaks it even in the case of flat metrics. 

If one restricts the state sum to only these flat metrics, by using hypercuboids, then it could be shown that the RG flow has an UV-attractive fixed point, on which vertex displacement symmetry is roughly restored. Since one only considers flat configurations, only the coupling constant $\alpha$ plays a role. Depending on the boundary state, the fixed point lies around $\alpha\approx 0.63$ \cite{Bahr:2017klw}. In the following, we extend the RG flow to frusta geometries which also allow for curvature. 

\begin{figure}[h]
	\includegraphics[scale=0.85]{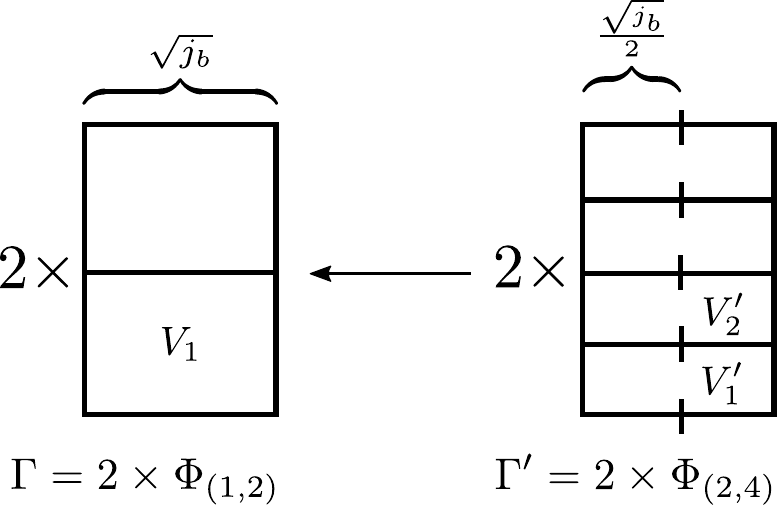}
	\caption{Coarse graining step used to generate the one-dimensional flow in the isochoric setting i.e., keeping the total 4-volume fixed.}
	\label{fig:CF1}
\end{figure}

We consider the coarse-graining step of $\Gamma=2\times\Phi_{(1,2)}$ into $\Gamma'=2\times\Phi_{(2,4)}$ which are respectively discretizations with $n_{\Gamma}=1^3\times2\times2=4$ and $n_{\Gamma'}=2^3\times4\times2=64$ vertices (Figure \ref{fig:CF1}). 
The lattice is doubled in one of the spatial directions, so that the amplitude is always positive. The initial and final boundary spins are fixed and equal $j_i=j_f=j_b$. 

The RG flow is then evaluated in the isochoric regime, i.e.~summing over all configurations which have identical total $4$-volume $V_\text{tot}$. This is achieved by performing a transformation of the integral over spins $(j_n,k_m)$ to an integral over $(j_n,V_m)$, with the $4$-volumes $V_m$ of a vertex at time-step $m$. This adds a Jacobian determinant to the integration, after which the total volume is fixed by including a $\delta(\sum_mV_m-V_\text{tot})$ into the integral, which allows to express one of the volumes by the others and $V_\text{tot}$.
For the coarse lattice $\Gamma$ this results in two variables $j_1, V_1$, while for the fine lattice $\Gamma'$ one has six variables $j'_n,V'_n$, with $n=1,2,3$. 

We use the amplitude \eqref{drevampl} and equation \eqref{expvalG} to compute the expectation values of an observable corresponding to the fluctuation of half of the volume, i.e.
\begin{eqnarray}\label{Eq:Isochoric_coarse}
	\langle O_{\Gamma}\rangle_{\Gamma}\;&\equiv& \;\langle(V_1-V_{\text{tot}}/2)^2\rangle,\\[5pt]\label{Eq:Isochoric_fine}
	\quad \langle O_{\Gamma'}\rangle_{\Gamma'}\;&\equiv&\; \langle (V'_1+V'_2-V_{\text{tot}}/2)^2\rangle.
\end{eqnarray}

\noindent To compare to the computation in \cite{Bahr:2017klw}, we fix $1/G=1.5$, $\Lambda=0.1$, and consider the amplitude depending only on the coupling constant $\alpha$. For a given $\alpha'$ on the fine lattice, we compute the fine observable (\ref{Eq:Isochoric_fine}), and look for the value $\alpha$ on the coarse lattice, which leads to the same value for (\ref{Eq:Isochoric_coarse}), i.e.~the RG flow $\alpha'\to \alpha$ is given by the condition
\begin{eqnarray}
	\langle \mathcal{O}_{\Gamma} \rangle^{\alpha}_{\Gamma}\;\stackrel{!}{=}\;\langle \mathcal{O}_{\Gamma'} \rangle^{\alpha'}_{\Gamma'}.
\end{eqnarray}

\begin{figure}
	\begin{center}
		\includegraphics[width=0.45 \textwidth]{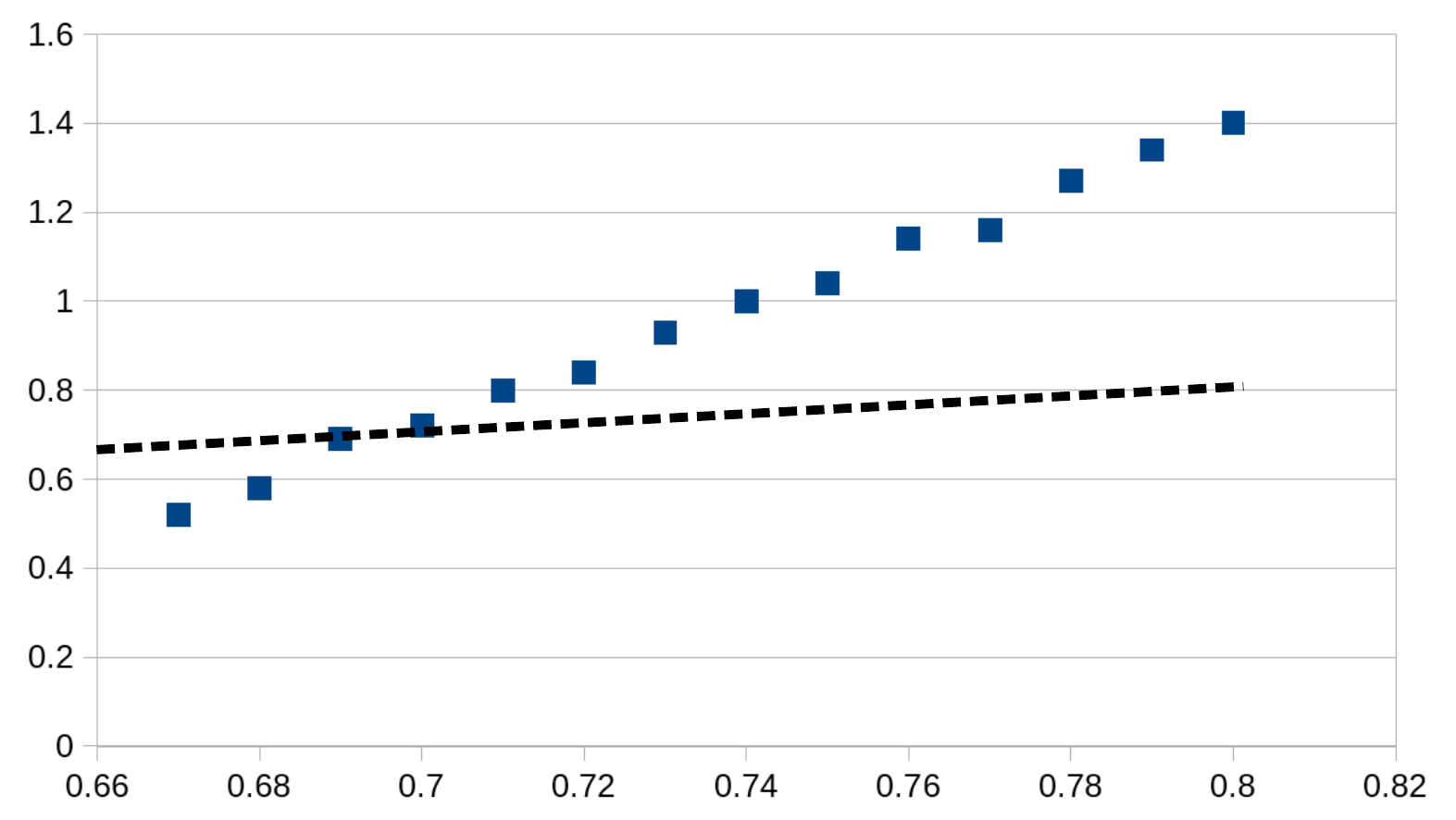}
		\caption{\label{Fig:Iso_RGFixedPoint}
			RG flow $\alpha\to\alpha'$ in the isochoric case. The intersection with the dashed line ($\alpha\alpha'$) lies at about $\alpha\approx 0.69$, while the other coupling constants are fixed to $1/G=1.5$, $\gamma=\frac{1}{2}$, and $\Lambda=0.1$. }
	\end{center}
\end{figure}

\noindent The result can be seen in Figure \ref{Fig:Iso_RGFixedPoint}. 
The intersection with the line of $\alpha=\alpha'$ lies at about
\begin{eqnarray}
	\alpha^*\;\approx\;0.69,
\end{eqnarray}
\noindent which marks an unstable (i.e.~UV-attractive) fixed point of this flow. This value is slightly above the one found in \cite{Bahr:2016hwc}, but only differs by about 10\%. 
\begin{figure}
	\begin{center}
		\includegraphics[width=0.45 \textwidth]{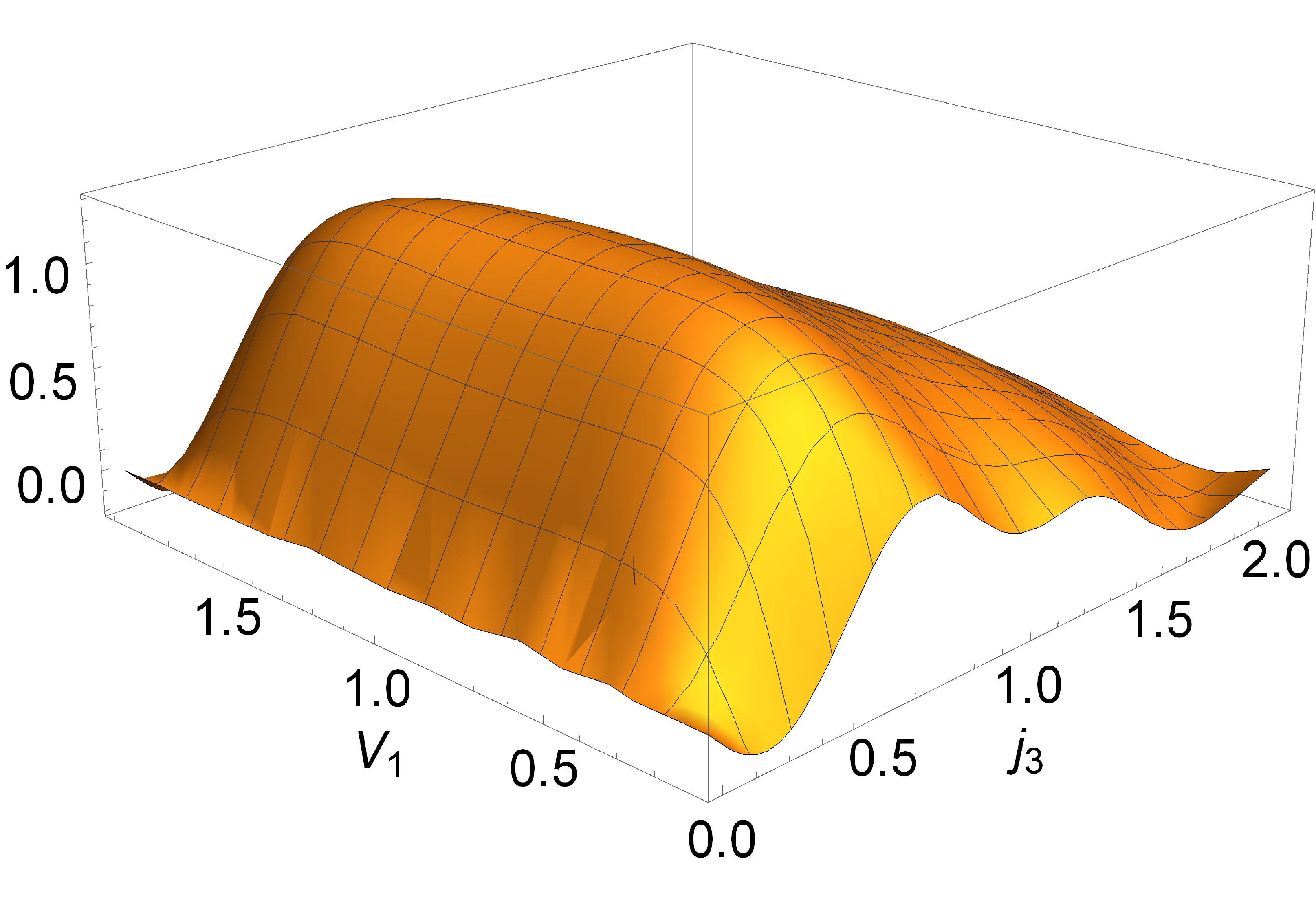}
		\caption{\label{Fig:Iso_IntegrandAtFixedPoint}
			Path integrand $\widehat{\mathcal{A}_1}\cdot\widehat{\mathcal{A}_2}$ for the coarse lattice at $\alpha=\alpha^*$, depending on the two variables $j_1$, $V_1$. The plateau indicates the presence of vertex displacement symmetry.}
	\end{center}
\end{figure}

A plot of the path integrand for the coarse lattice (depending on the two free variables $j_1$, $V_1$) is depicted in Figure \ref{Fig:Iso_IntegrandAtFixedPoint}. It can be seen that for $\alpha$ at the fixed point, there is a plateau in the integrand, indicating that some symmetry among the variables is approximately realised in the path integral. This can be regarded as some vertex displacement symmetry. It should be noted, however, that in this case the connection to the diffeomorphisms is much less clear, due to the presence of non-trivial deficit angles. 
Numerical tests show indeed that the plateau depicted in Figure \ref{Fig:Iso_IntegrandAtFixedPoint} vanishes, as soon as one moves $\alpha$ away from the fixed point $\alpha^*$. All of this is in agreement with what has been found previously in the case of hypercuboids \cite{Bahr:2016hwc, Bahr:2017klw}. 

\subsection{The isotemporal gauge}  
Let us now go beyond the one dimensional analysis and generate higher dimensional RG flow diagrams. In fact, the theory is also defined by the parameters $G$ and $\Lambda$. 
We first look at a two dimensional flow in the space $(\Lambda,G)$ while keeping fixed the value of $\alpha$. Such analysis reveals a partial information being a projection of the three dimensional flow. Nonetheless we will see that it carries the traces of non trivial regions.
We then extend this result to the entire parameter space generating a more detailed flow diagram in the space  $(\alpha,G,\Lambda)$. As we will show, the flow has a fixed point with one repulsive and two attractive directions.

Here we relax the constraint which keeps fixed the total 4-volume and instead we fix the total height $H$. Furthermore, we  work in an isotemporal gauge, i.e.~we demand that all the hyperfrusta in a given discretization have the same height. As an example, the slot $(1,\mathcal{Y})$ of Fig.\ref{fig:Discretizations} is now interpreted as the transition of a single cube into the same cube in $\mathcal{Y}$ time steps which is represented by a chain of $\mathcal{Y}$ hyperfrusta of same height $H/\mathcal{Y}$.

\begin{figure}[h]
	\includegraphics[scale=0.85]{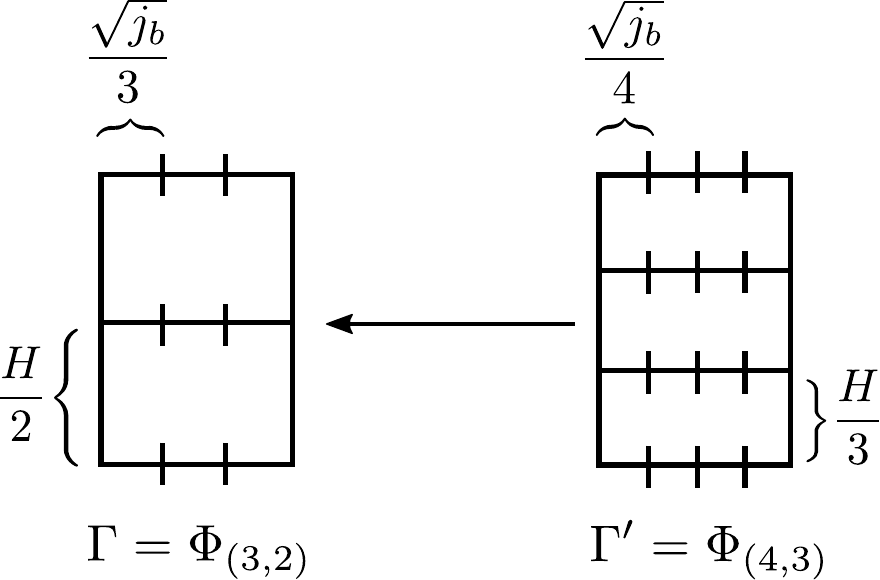}
	\caption{Coarse graining step used to generate the two- and three-dimensional flows in the isotemporal gauge, i.e. keeping fixed the height of the vertices in each discretization and imposing the total height $H$ to be fixed.}
	\label{fig:CF23}
\end{figure}

In our analysis we consider the case of $\Gamma=\Phi_{(3,2)}$ and $\Gamma'=\Phi_{(4,3)}$ which respectively correspond to discretizations of $\Phi$ in terms of $n_{\Gamma}=3^3\times 2=54$ and $n_{\Gamma'}=4^3\times 3=192$ hyperfrusta (Figure \ref{fig:CF23}).
We also choose a fiducial set of boundary conditions $j_i=j_f=1$ and we fix $H=6$. 

Let us note that the total amplitude of $\Gamma'$ is always positive being given as a product of an even number of identical dressed vertex amplitudes for each time step.
The coarse lattice $\Gamma$ has instead an odd number of vertices contributing to each time step. However, thanks to the symmetry $\widehat{\mathcal{A}}_{v}(j_n,j_{n+1},k_n)=\widehat{\mathcal{A}}_{v}(j_{n+1},j_n,k_n)$ of \eqref{drevampl}, the chosen boundary conditions and the isotemporal gauge setting guarantee the positivity of the total amplitude as both time steps carry the same amplitude. 

In the large spin limit the partition functions associated to these two systems are respectively 
\begin{equation}
	\begin{split}
		Z_{\Gamma}=\int& \mathrm{d}j_1 \mathrm{d}k_1 \mathrm{d}k_2 \ \widehat{\mathcal{A}}_v(\frac{j_i}{9},j_1,k_1)^{27} \  \widehat{\mathcal{A}}_v(j_1,\frac{j_f}{9},k_2)^{27},\\
		Z_{\Gamma'}=\int& \mathrm{d}j'_1 \mathrm{d}j'_2 \mathrm{d}k'_1 \mathrm{d}k'_2 \mathrm{d}k'_3 \ \widehat{\mathcal{A}}_v^{64}(\frac{j_i}{16},j'_1,k'_1) \\
		& \times  \ \widehat{\mathcal{A}}_v^{64}(j'_1,j'_2,k'_2) \ \widehat{\mathcal{A}}_v^{64}(j'_2,\frac{j_f}{16},k'_3),
	\end{split}
\end{equation}
where $j_1,j'_1,j'_2$ are internal space-like spins associated to square areas while $k_1,k_2,k'_1,k'_2,k'_3$ are internal `time-like' spins associated to trapezoidal faces. 
To implement the isotemporal gauge we first perform a change of variables
\begin{equation}
	\begin{split}
		& k_1\rightarrow H_1, \qquad k_2\rightarrow H_2, \qquad \\
		& k'_1\rightarrow H'_1 , \qquad  k'_2\rightarrow H'_2,  \qquad k'_3\rightarrow H'_3.
	\end{split}
\end{equation}
Each of these substitutions generates a Jacobian factor. For an hyperfrustum $\mathrm{F}_n(j_n,j_{n+1},k_n)$ the Jacobian $J\equiv J(j_n,j_{n+1},H_n) $ reads
\begin{equation}
	\begin{split}
		J&=\frac{\partial H_n(j_n,j_{n+1},k_n)}{\partial k_n}\\
		& = \frac{H_n(\sqrt{j_n}+\sqrt{j_{n+1}})^2}{\sqrt{4 H_n(\sqrt{j_n}+\sqrt{j_{n+1}})^2+2(j_n-j_{n+1})^2}},
	\end{split}
\end{equation}
and refers to the change of variables $k_n \rightarrow H_n$, the height $H_n$ being given in terms of \eqref{short} by
\begin{equation}\label{heighthyperfr}
	H_n= \frac{2k_n}{\sqrt{j_{n+1}}+\sqrt{j_n}}  \ K.
\end{equation}
As a second step we insert in the coarse and fine partition functions respectively 
\begin{equation}
	\begin{split}
		& \delta(H-H_1-H_2)\delta(H_1-H_2), \\
		\\
		& \delta(H-H'_1 -H'_2- H'_3)\delta(H'_1-H'_2)\delta(H'_1-H'_3).
	\end{split}
\end{equation}
The partition functions then become
\begin{equation}
	\begin{split}
		Z_{\Gamma}=\int& \mathrm{d}j_1 \ {\mathcal{A}}_{\Gamma},\\
		Z_{\Gamma'}=\int& \mathrm{d}j'_1 \mathrm{d}j'_2 \ {\mathcal{A}}_{\Gamma'},
	\end{split}
\end{equation}
where we have defined 
\begin{equation}
	\begin{split}
		{\mathcal{A}}_{\Gamma} &= J\Big(\frac{j_i}{9},j_1,\frac{H}{2}\Big) \  J\Big(j_1,\frac{j_i}{9},\frac{H}{2}\Big)\\ & \times {\widehat{\mathcal{A}}}^{27}\Big(\frac{j_i}{9},j_1,k_1\Big) \  {\widehat{\mathcal{A}}}^{27}\Big(j_1,\frac{j_i}{9},k_2\Big),\\
		{\mathcal{A}}_{\Gamma'} &= J\Big(\frac{j_i}{16},j'_1,\frac{H}{3}\Big)  \ J\Big(j'_1,j'_2,\frac{H}{3}\Big) J\Big(j'_2,\frac{j_f}{16},\frac{H}{3}\Big)\\
		& \times  {\widehat{\mathcal{A}}}^{64}\Big(\frac{j_i}{16},j'_1,k'_1\Big) \  {\widehat{\mathcal{A}}}^{64}\Big(j'_1,j'_2,k'_2\Big){\widehat{\mathcal{A}}}^{64}\Big(j'_2,\frac{j_f}{16},k'_3\Big).
	\end{split}
\end{equation}
The `time-like' spins in the expressions above must be understood as functions $k_n\equiv k_n(j_n,j_{n+1},H_n)$.

Thus, in the coarse case we remain with a system with a single d.o.f. given by the intermediate spatial spin $j_1\in [0,\infty]$. In the fine case there are two d.o.f.~corresponding to the two intermediate spins $j'_1,j'_2\in [0,\infty]$. 

We evaluate expectation values of $n$ observables  $\mathcal{O}^{(n)}$ by numerically integrating over these variables 
\begin{equation}
	\begin{split}
		\langle\mathcal{O}^{(n)}_{\Gamma}\rangle_{\Gamma}^{g}=\frac{1}{Z_{\Gamma}}\int& \mathrm{d}j_1 \ \mathcal{O}^{(n)}_{\Gamma} \ {\mathcal{A}}_{\Gamma},\\
		\langle\mathcal{O}^{(n)}_{\Gamma'}\rangle_{\Gamma'}^{g'}=\frac{1}{Z_{\Gamma'}}\int& \mathrm{d}j'_1  \mathrm{d}j'_2 \ \mathcal{O}^{(n)}_{\Gamma'} \ {\mathcal{A}}_{\Gamma'},
	\end{split}
\end{equation}
where $g=(\alpha,G,\Lambda)$ and $g'=(\alpha',G',\Lambda')$ are sets of parameters defining the theory. 
Observables should be cylindrically consistent, written as:
\begin{equation}\label{cyl}
	\langle\mathcal{O}^{(n)}_{\Gamma}\rangle_{\Gamma}^{g}=
	\langle\mathcal{O}^{(n)}_{\Gamma'}\rangle_{\Gamma'}^{g'} \qquad \forall n.
\end{equation} 
This can be seen as equations for the coupling constants $g$, $g'$, which defines the RG flow equation for our model. 
In fact, if one can solve it, for any point $g'$ the equation returns a point $g$ and we can connect them with an arrow $\mathrm{A}^{g'\rightarrow g}$ to draw the flow in the parameter space \footnote{In analogy with the RG flows generated in the Asymptotic Safety scheme, where the arrows point from high to low energy, here the arrows start at $g'$ associated to the fine observables, and point at $g$ which is related to coarse observables. We recall that, in our context of background independent renormalization, there are no continuous labels tracing the energy scale. Instead, the shift of resolution happens in discrete steps and is associated to a change of discretization. This also equates to a change in the number of degrees of freedom that we keep when describing a physical process. 
	Thus, in a `Wilsonian' sense, the refinement of a discretization can be interpreted as a shift towards high energy regimes.}.
The existence of an exact solution to equation \eqref{cyl} depends on many factors. We already discussed the relevance of the choice of $\Gamma$ and $\Gamma'$ as well as the various approximations that may spoil the solution. A further technical obstacle is represented by the fact that the solution of \eqref{cyl} would require the knowledge of the values $\langle\mathcal{O}^{(n)}_{\Gamma}\rangle_{\Gamma}$ and  $\langle\mathcal{O}^{(n)}_{\Gamma'}\rangle_{\Gamma'}$ in all the points of the parameter space.
However, in our case these observables are evaluated numerically for every couple $(g,g')$. Therefore we must consider a finite number of points in the parameter space in order to perform a finite number of integrations. The solution of the flow equation is then approximated whereas for a point $g'$ we cannot access all the points in its neighbourhood with infinite accuracy and, consequently, the point $g$ cannot be defined exactly. Note that here we also assume implicitly that the RG flow makes only small steps in the coupling constants. While this can be expected to hold near a fixed point, in general it might not be true, increasing the error of our RG computation. 

In the light of this observations we impose the cylindrical consistency condition in a weak form
\begin{equation}\label{weakcyl}
	\Delta^{g,g'}_{\Gamma, \Gamma'}\equiv \sum_n|\langle\mathcal{O}^{(n)}_{\Gamma}\rangle_{\Gamma}^{g}-\langle\mathcal{O}^{(n)}_{\Gamma'}\rangle_{\Gamma'}^{g'}|\stackrel{!}{=}\mathrm{min}.
\end{equation}
Our plan consists in considering an adequate number of points in a `large' region of the parameter space, determine the flow accordingly to the weak cylindrical consistency condition \eqref{weakcyl} and finally, for each arrow $\mathrm{A}^{g'\rightarrow g}$, check how small is the relative error
\begin{equation}\label{relerr}
	R^{g,g'}_{\Gamma, \Gamma'}\equiv\frac{\Delta^{g,g'}_{\Gamma, \Gamma'}}{\overline{\mathcal{O}}^{g,g'}_{\Gamma, \Gamma'}},
\end{equation} 
being
\begin{equation}
	\overline{\mathcal{O}}^{g,g'}_{\Gamma, \Gamma'}\equiv\sum_n\Bigg|\frac{\langle\mathcal{O}^{(n)}_{\Gamma}\rangle_{\Gamma}^{g}+\langle\mathcal{O}^{(n)}_{\Gamma'}\rangle_{\Gamma'}^{g'}}{2}\Bigg|.
\end{equation}

Regions of parameter space interesting for the RG flow (e.g.~since one expects a fixed point there) can be studied with higher accuracy by zooming further into that region.
During our analysis we encountered many regions of the parameter space where the cylindrical consistency condition is in fact violated and the RG flow cannot be trusted. We concentrate on those regions where cylindrical consistency is satisfied up to only small errors. 

\subsection{Two dimensional isotemporal RG flow}
Let us look at the projection of the RG flow on the two dimensional parameter space $(\Lambda,G)$. To do so we fix the value of $\alpha=0.68$. 
We recall that the choice of $\alpha$ influences the convergence of the path integral. 
In particular, the chosen value for $\alpha$ favours small spins. This allows us to set an upper spin cutoff during the Monte Carlo integrations so that the results will be independent of it. Furthermore, this value of $\alpha$ stands out in our analysis as a point where an interesting and consistent dynamics is expected to take place, as indicated by our earlier investigation in section \ref{Sec:1DFlow}. 

\begin{figure*}[ht]
	\hspace*{-1cm}
	\begin{minipage}{\dimexpr\linewidth-0.50cm\relax}%
		\begin{minipage}{0.5cm}
			\vspace*{-8.8cm}
			\rotatebox{90}{G}
		\end{minipage}%
		\includegraphics[trim=0 0 15 14,clip,scale=0.5]{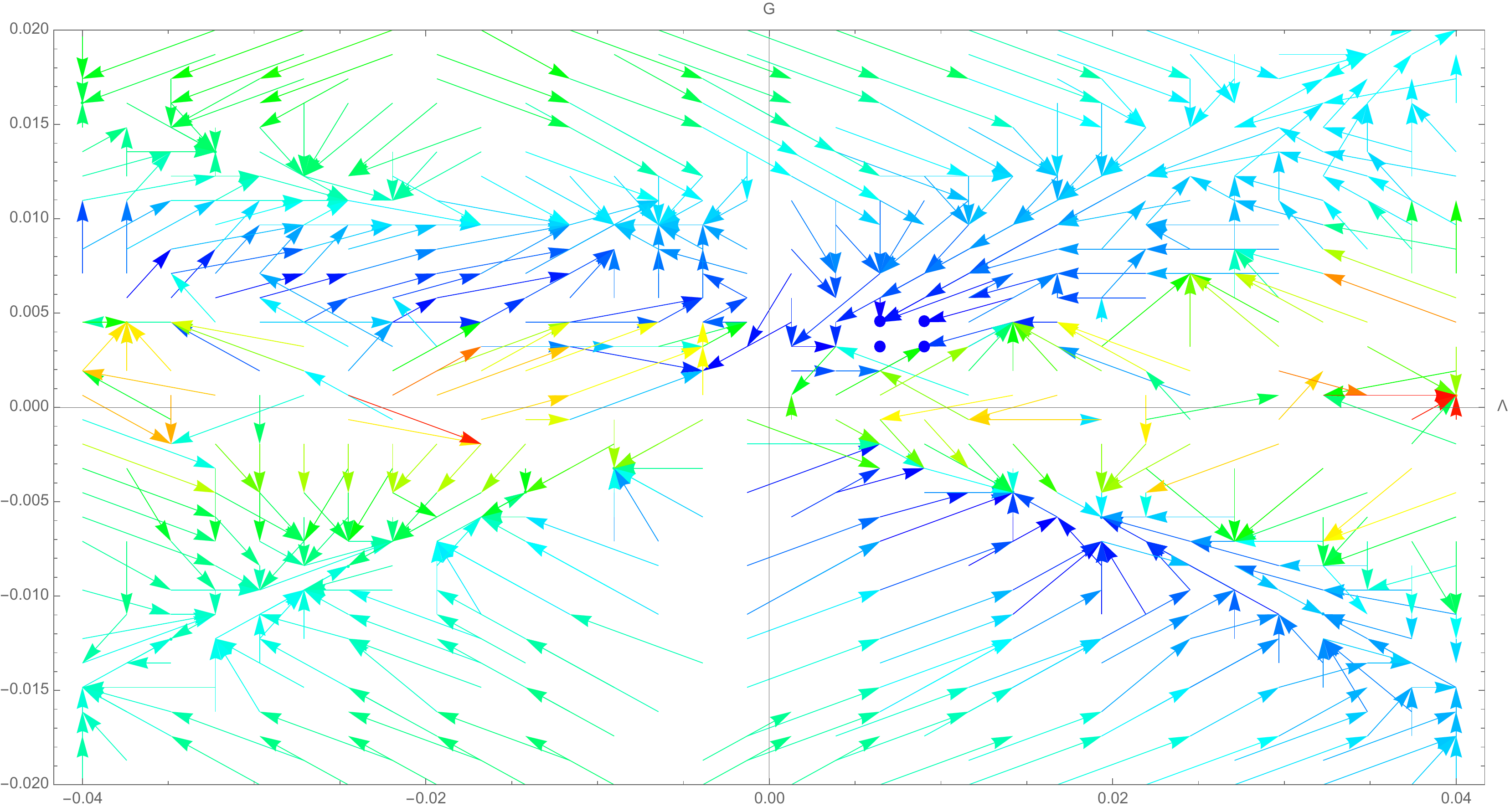}
		\vspace*{0.1cm}\hspace*{1.0cm}$\Lambda$
	\end{minipage}%
	\caption{RG flow with cylindrical consistency condition maximally and minimally violated with the respective relative errors  $R_{red}=4.0675$, $R_{blue}=0.0169$.}
	\label{fig:2Da}
\end{figure*}

In order to draw a flow diagram we proceed as follows 
\begin{itemize}
	\item Select a domain in the parameter space $(\Lambda, G)$ and identify $n=32\times 32=1024$ points homogeneously distributed in this domain.
	\item In each point of the domain evaluate numerically the coarse and fine expectation values of three operators: 
	\begin{enumerate}
		\item the 3-volume at middle height $\langle\mathcal{O}^{(1)}\rangle\equiv\langle V_3 \rangle$.
		\item its variance $\langle\mathcal{O}^{(2)}\rangle \equiv \langle V_3^2 \rangle- \langle V_3\rangle^2$.
		\item the total 4-volume $\langle\mathcal{O}^{(3)}\rangle\equiv\langle V_4 \rangle$. 
	\end{enumerate}
	\item Starting from each $g'=(\Lambda', G')$ draw an arrow $\mathrm{A}^{g'\rightarrow g}$ pointing at $g=(\Lambda, G)$ such that, following the notation of \eqref{weakcyl}, the distance $\Delta^{g^{\star},g'}_{\Gamma, \Gamma'}$
	is minimal for $g^{\star}=g$,  where $g^{\star}$ is a point in the selected domain. This defines an RG flow diagram.\footnote{In the first plots we fix a maximum length for the arrows since we are interested in getting an idea about where to zoom next to satisfy equation \eqref{weakcyl} best. Later, when we are in a region that we can trust, we will allow the arrows to have any length. }
	\item Assign a color to the arrows depending on the value of the relative errors $R^{g,g'}_{\Gamma, \Gamma'}$,
	where we have used the notation as in \eqref{relerr}. Namely, draw in red the arrow that violates the most the cylindrical consistency condition \eqref{weakcyl} (w.r.t. the other arrows in the plot). On the contrary, color in blue the one which satisfies best the condition. Report the corresponding values $R_{red}$ and $R_{blue}$ of the relative errors. According to the above classification, draw the other arrows in a tonal progression from red to blue.
\end{itemize}
The resulting RG flow in the region $\Lambda=(-0.04, 0.04)$ and $G=(-0.02,0.02)$ is shown in Figure \ref{fig:2Da}.

\begin{figure*}[!ht]
	\hspace*{-1cm}
	\begin{minipage}{\dimexpr\linewidth-0.50cm\relax}%
		\begin{minipage}{0.5cm}
			\vspace*{-8.8cm}
			\rotatebox{90}{G}
		\end{minipage}%
		\includegraphics[trim=0 0 15 14,clip,scale=0.5]{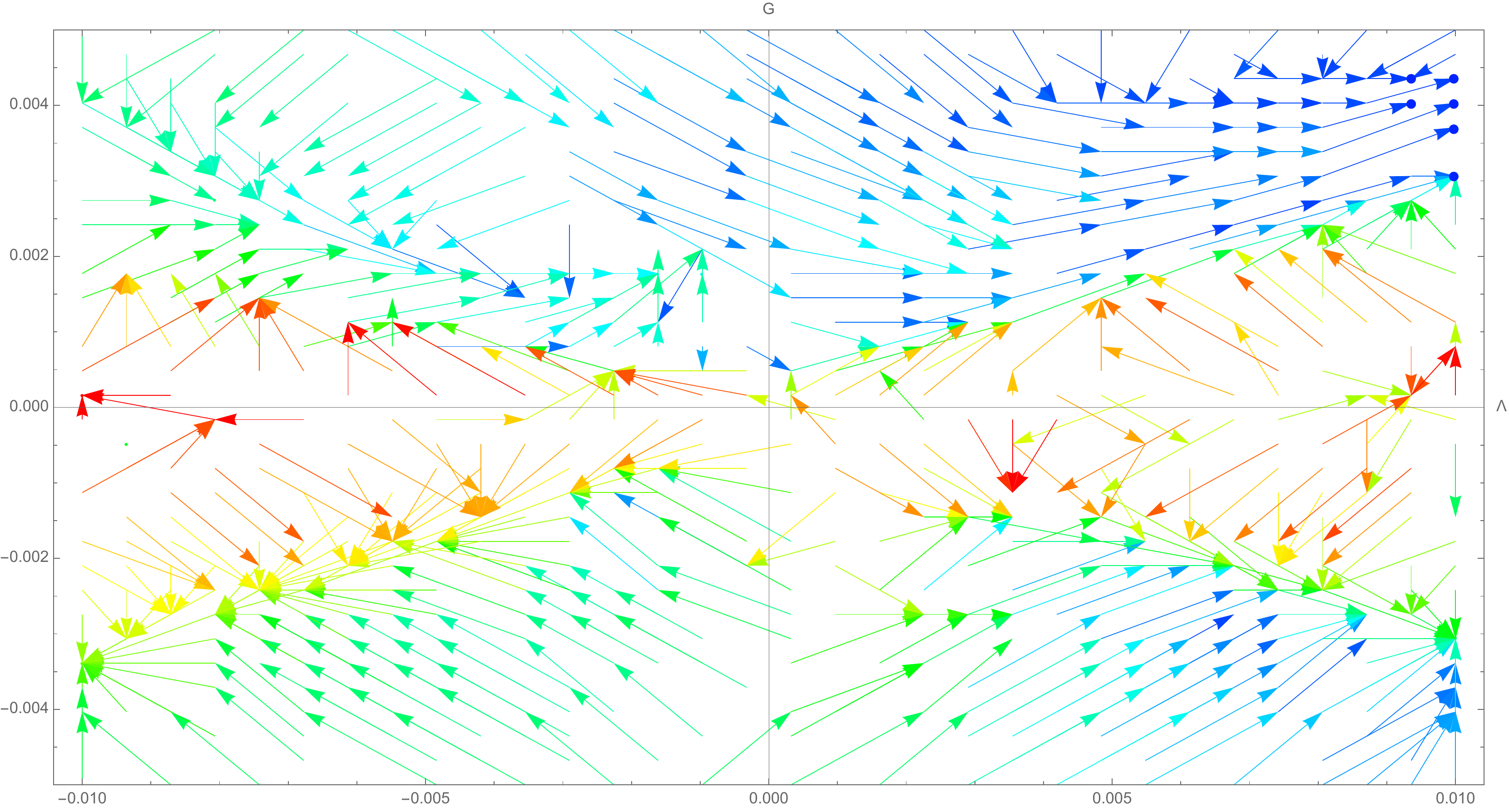}
		\vspace*{0.1cm}\hspace*{1.0cm}$\Lambda$
	\end{minipage}%
	\caption{RG flow with cylindrical consistency condition maximally and minimally violated with the respective relative errors  $R_{red}=4.0675$, $R_{blue}=0.0169$.}
	\label{fig:2Db}
\end{figure*}
\clearpage

As the relative errors suggest, at the analysed resolution the flow is hardly reliable in some regions. Still we notice that the arrows drawn in dark blue have a small relative error $R\sim0.017$. Most notably, those in the first quadrant, close to $(\Lambda,G)=(0,0)$, show an interesting behaviour where  they have a vanishing length (represented by dots). This is exactly what we would expect to happen at a fixed point. Let us then zoom into such region. The result for $\Lambda=(-0.01, 0.01)$ and $G=(-0.004,0.004)$ is shown in Figure \ref{fig:2Db}.



A first clear observation is that the overall relative errors have improved, reaching a top precision $R\sim0.008$. In an angular region around $G=0$ the flow is still unreliable. However, in agreement with the interesting region (blue arrows), the relative errors are fairly small and the flow shows a more coherent behavior. In particular there are still some arrows with null distance. We then want to zoom further into the top right region of Figure \ref{fig:2Db}. We do so by also unlocking the parameter $\alpha$ and let it vary slightly around $\alpha=0.68$.

\subsection{Three dimensional isotemporal RG flow}
\begin{figure*}[!htb]
	\centering
	\quad \includegraphics[scale=0.6]{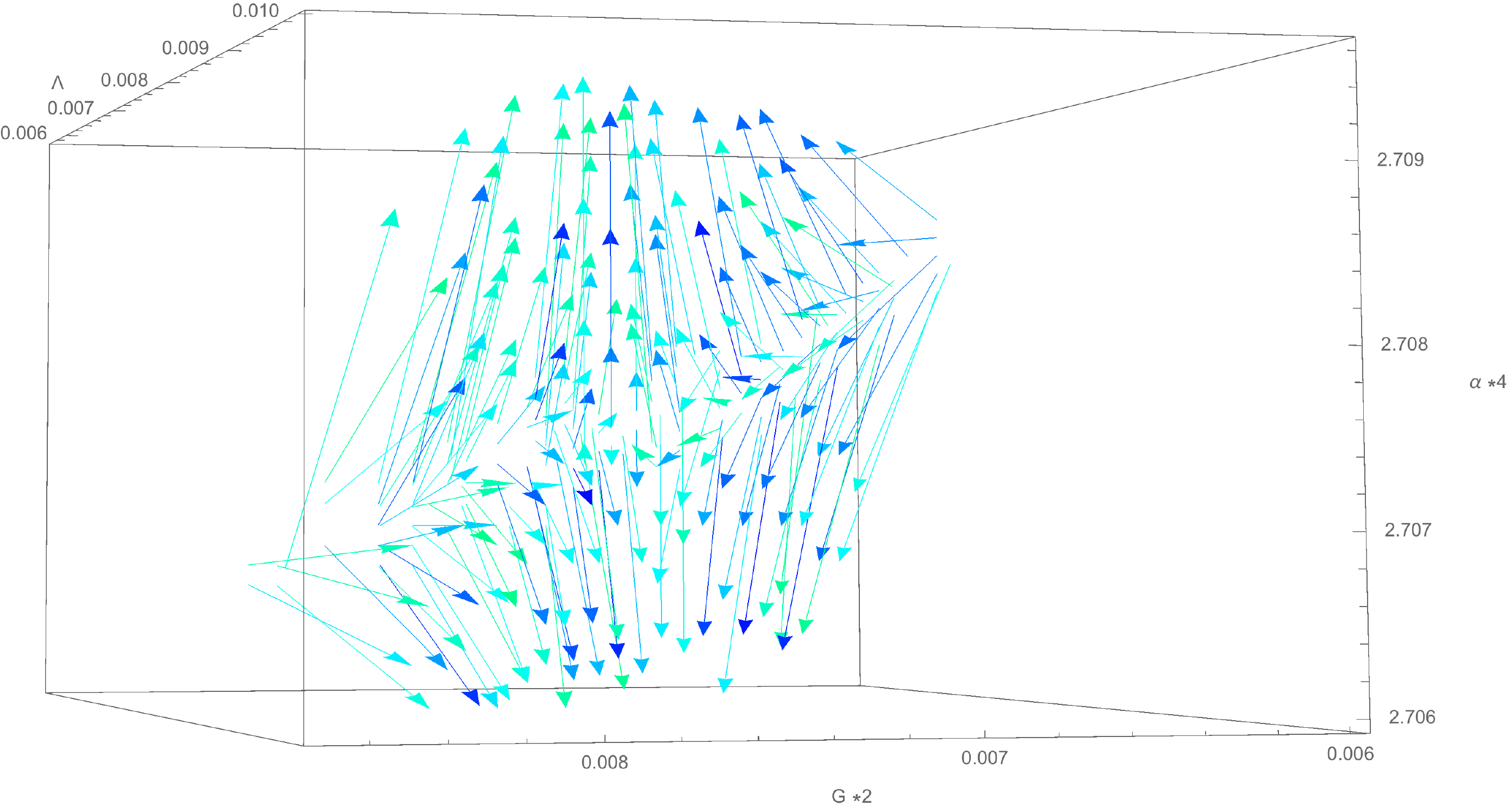}
	\caption{Three dimensional RG flow with cylindrical consistency condition maximally and minimally violated with the respective relative errors  $R_{green}=0.004$, $R_{blue}=0.00017$.}
	\label{fig:3D}
\end{figure*}

Using the same strategy as in the two dimensional case, it is possible to generate an RG flow in the space defined by the three coupling constants $(\alpha, G, \Lambda)$. Figure \ref{fig:3D} shows the RG flow in the region $\Lambda=(0.006, 0.01)$, $G=(0.003,0.0045)$ and $\alpha=(0.6765, 0.6775)$ in which we have selected $32\times32\times32$ points. All the arrows in the plot satisfy  the cylindrical consistency condition to high precision, the smallest relative error being $R\sim0.00017$ \footnote{ For practical graphical reasons, we only draw the most reliable arrows in blue and green. }.

Remarkably, there is the indication of a fixed point within the center of this region, showing one repulsive and two attractive directions. At this order of precision, both the relevant (repulsive) and irrelevant (attractive) directions seem to be associated with linear combinations of all three parameters. A better precision can be reached by further zooming.  Our research suggest that this is a rare point of the parameter space. Whether this point is unique needs further analysis.

\section{Expanding and contracting universes}\label{Sec:ExpandingContracting}

We now investigate the dynamics described by the amplitudes, in order to gain an insight into the interpretation of the RG flow. 

Frusta geometries are geared towards studying cosmological transitions. The spatial cubes essentially encode the scale factor $a$ of the universe at a certain `time step', and the `time-like' frusta mediate between spatial cubes of different size\footnote{The cuboid intertwiners we use are sharply peaked on the cuboid shape, yet they are undetermined in the extrinsic curvature, i.e. how the 3D cubes are embedded in a 4D geometry. In this sense the states are sharply peaked in $a$, but $\dot{a}$ is maximally uncertain.}. Naturally the question arises which configurations are preferred in the path integral given by the EPRL amplitudes. In particular we intend to examine how the parameters of the model, e.g. the cosmological constant $\Lambda$, influence the dynamics and whether familiar features of the classical theory emerge as well. In the case of our simple model this could be whether the universes expansion is accelerating or slowing down, depending on the sign of the  cosmological constant.

\begin{figure*}
	\includegraphics[scale=0.75]{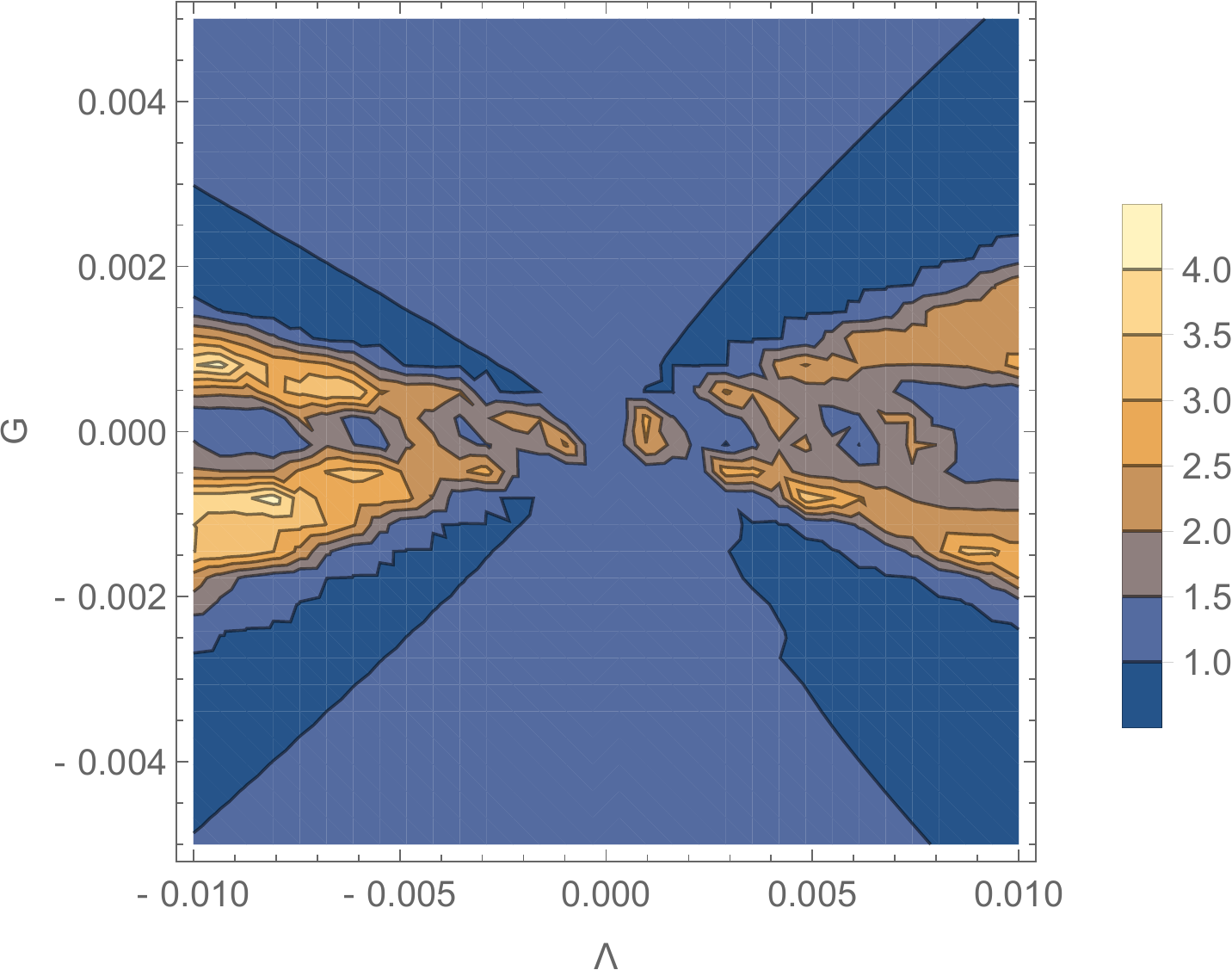}
	\caption{\label{fig:3D_volume_coarse}
		Expectation value for the 3D volume of the intermediate spin, $\langle j^{\frac{3}{2}}\rangle$, for $\alpha= 676855$. This is the case of the coarse transition with 54 hyperfrusta.}
\end{figure*}

To this end, we study again the expectation values of observables that we have used before to define and compute a renormalization group flow. More precisely, we consider the 3D volume for the coarse transition investigated before, as it essentially gives the intermediate scale factor between an initial and final state of the same size. Furthermore, studying an observable used for the renormalization group flow in more detail may reveal a few insights as to the form of the flow. We show its expectation value in fig. \ref{fig:3D_volume_coarse}.

As a first striking feature, we recognize the `X'-shape in the values of the observables similar to the 2D scans of the renormalization group flow. Inside this region, the 3D volume fluctuates significantly and can reach quite high values. These peaks appear to be slightly larger for negative cosmological constant, but there also exist regions for positive $\Lambda$, in which the intermediate 3D volume is significantly larger compared to the initial / final state. Note that this is also the region in which the cylindrical consistency conditions for the observables of the RG flow are strongly violated, which implies that a similar behaviour does not exist in a similar region for the fine observable. Judging from the plot, this behaviour is due to the small size of $|G|$ and it appears to extend slightly as $|\Lambda|$ is increased. A possible explanation is that both parameters enhance the oscillatory behaviour of the integral, resulting in a highly fluctuating expectation value. 

Outside that region, more precisely for larger $|G|$, we observe a rather uniform behaviour, where the 3D volume is around or slightly larger than 1, which is also the volume at the initial and final slice.

There is only little dependence on the sign of the cosmological constant: For negative $\Lambda$, we observe a slightly larger intermediate 3D volume already for smaller $|G|$. Thus, $\Lambda < 0$ appears to favour a larger intermediate 3D volume compared to $\Lambda >0$, however in both cases we observe an intermediate volume that is larger than the initial and final one. Hence, we generically observe a transition in which the universe first expands and then contracts, or at most remains constant. A transition to a contracting and then expanding universe is not observed numerically.

Naturally, one would like to compare this behaviour to classical dynamics. However it is not clear to which discrete action we should compare our results to. In the vertex amplitude (\ref{drevampl}) several oscillating terms appear, containing different actions. While the cosine contains the (area) Regge action and a volume term times the cosmological constant, the other oscillating terms only contain the Regge action. Clearly the former term is the desired one, we will briefly compare our results to the classical, discrete dynamics.

Since we consider the transition for fixed heights, with $j_\text{in} = j_\text{fin} = 1$. There is one discrete equation of motion to solve depending on $\Lambda$\footnote{As $G$ is an overall constant, only $\Lambda$ determines the classical dynamics in the absence of matter.}. For $\Lambda=0$ the equations of motion are solved by $j=1$, so there is no expansion or contraction as one would expect. For $\Lambda > 0$ we find $j<1$ as the solution, while for $\Lambda < 0$ we find $j>1$. So we see a first contracting, then expanding universe for positive cosmological constant and the opposite for negative cosmological constant. Something similar can be seen in the continuum, where $\Lambda > 0$ implies $\ddot{a} > 0$. Hence in order to arrive at the same scale factor $a$ at a later time, the universe first contracts before expanding again. The behaviour is reversed as $\Lambda < 0$ implies $\ddot{a} < 0$.

It seems that the behaviour of the truncated SFM does not  reproduce the classical dynamics. Instead we usually see $\langle j^{\frac{3}{2}}\rangle > 1$, no matter the sign of the cosmological constant. Nevertheless, we do observe generically larger expectation values $\langle j^{\frac{3}{2}}\rangle > 1$ for negative $\Lambda$ compared to positive $\Lambda$. There are a few plausible explanations for these deviations: The vertex amplitude contains several oscillating functions, some contain the cosmological constant term, some do not. Moreover, the `proper' action appears in the cosine, which might lead to unwanted interference of different bulk solutions. Additionally, the whole spin foam does not oscillate with the sum of Regge actions assigned to hyperfrusta, as the cosine is not additive. Another possible deviation might stem from the face amplitudes, which favour small or large spins depending on the value of the parameter $\alpha$. If $\alpha$ is large, it puts emphasis on large spins, which generically results in larger expectations values for spins or volume etc.

A possibility to overcome the `cosine' problem would be to consider states which are not just peaked on the shape of cuboids or frusta, but which are also peaked in the extrinsic curvature. This would roughly correspond to prescribing both $a$ and $\dot{a}$ at the initial and final time. As a result, one of the two stationary and critical points in the asymptotic expansion might be suppressed, resulting in a quantum dynamics closer to its classical counterpart. We leave this for future research.

\section{Free theory}\label{Sec:FreeTheory}

In this chapter we consider the limit of the RG flow equations $G\to 0$ and $\Lambda\to 0$. This can be understood as the free theory, as the gravitational coupling $G$, which governs the strength of the perturbative interaction in the linearized theory, vanishes. It should be noted that, due to its non-perturbative nature, the EPRL-FK model does not exist for $G= 0$ ($\Lambda=0$ is no problem, though). We therefore approach this point in theory space asymptotically. 

Considering the RG step of a lattice with $4 \times 4\times 4 \times 3=192$ to one with $3\times 3\times 3 \times 2=54$ vertices, as described in section \ref{Sec:Approximations}. We compute the observables $V_3$ and $V_4$ for the isotemporal case, i.e.~when the time-steps are gauge-fixed, for $\Lambda=0$ in the asymptotic limit $1/G\to\infty$. The initial and final boundary spins are fixed to the same (but ultimately arbitrary) value $j_i=j_f=j$.

We first consider not the full EPRL-FK model, but only its proper vertex, where the amplitude as replaced simply by the exponential of the Regge action. In that case, we have that
\begin{eqnarray}\label{Eq:Free:StateSum54}
	Z_{54}\;=\;\int_0^{J_\text{max}}dj_1\;\Big(\hat{\mathcal{A}}\Big)^{54} \; ,
\end{eqnarray}

\noindent with
\begin{eqnarray*}
	\hat{\mathcal{A}}\;=\;F(j_1)\frac{e^{54 i/G\,S_R}}{|D|} \; ,
\end{eqnarray*} 

\noindent where $D$ is the Hessian determinant, and $S_R=S_R(j_1,j, H)$ is the Regge action for one hyperfrustum with initial/final spin $j$, intermediate spin $j_1$, and height $H$. Also, $F(j_1)$ is a function depending on $j_1$ (and $j$ and $H$), which are given by a collection of face- and edge-amplitudes. 

To evaluate (\ref{Eq:Free:StateSum54}) in the limit $1/G\to\infty$, we can perform a stationary phase approximation. For this we simply observe that the condition
\begin{eqnarray}\label{Eq:Free:StationaryEquation}
	\frac{\partial S_R(j,j_1,k(j,j_1,H))}{\partial j_1}\;=\;0
\end{eqnarray}

\noindent has only $j_1=j$ as solution. To compute expectation values, we perform the same calculation, but include another function $O(j,H,j_1)$ (in our case $V_3$ and $V_4$) into the integral, which we evaluate at the respective stationary point as well. We can immediately conclude that
\begin{eqnarray}
	\langle V_3\rangle_{54}^{G\to 0,\Lambda=0}\;=\;27j^{\frac{3}{2}},\quad
	\langle V_4\rangle_{54}^{G\to 0,\Lambda=0}\;=\;54H\,j^{\frac{3}{2}}.
\end{eqnarray}

\noindent The computation for $Z_{192}$ is only slightly more complicated. We have
\begin{equation}
	Z_{192}\;=\;\int dj'_1dj'_2 \left(\hat{\mathcal{A}}_1\hat{\mathcal{A}}_2\hat{\mathcal{A}}_3 \right)^{64},
\end{equation}

\noindent where the $\hat{\mathcal{A}}_i$, $i=1,2,3$ denote the vertex amplitudes for the $i$-th time step. We get
\begin{eqnarray}
	\hat{\mathcal{A}}_1\hat{\mathcal{A}}_2\hat{\mathcal{A}}_3 \;=\;\frac{e^{i/G\,\Big(S_1+S_2+S_3\Big)}}{|D_1D_2D_3|} \; ,
\end{eqnarray}

\noindent with the Regge actions $S_i$ for the $i$-th time step, and $D_i$ the corresponding Hessian determinant. The variables for these are $j'_1$ and $j'_2$, and one can show that, again, the only solution to
\begin{eqnarray}
	\frac{\partial}{\partial j'_1}(S_1+S_2+S_3)\;=\;\frac{\partial}{\partial j'_2}(S_1+S_2+S_3)\;=\;0
\end{eqnarray}

\noindent is $j'_1=j'_2=j'$. This immediately leads to
\begin{eqnarray*}
	\langle V_3\rangle_{192}^{G\to 0,\Lambda=0}\;=\;64(j')^{\frac{3}{2}},\quad
	\langle V_4\rangle_{192}^{G\to 0,\Lambda=0}\;=\;192H'\,(j')^{\frac{3}{2}}.
\end{eqnarray*}

\noindent With $H'=\frac{2}{3}H$ and $\sqrt{j'}=\frac{3}{4}\sqrt{j}$, we can conclude that
\begin{eqnarray*}
	\langle V_3\rangle_{192}^{G\to 0,\Lambda=0}\;&=&\;\langle V_3\rangle_{54}^{G\to 0,\Lambda=0},\\[5pt]
	\langle V_4\rangle_{192}^{G\to 0,\Lambda=0}\;&=&\;\langle V_4\rangle_{54}^{G\to 0,\Lambda=0}.
\end{eqnarray*}

\noindent This demonstrates that the point $G=0$, $\Lambda=0$ is a fixed point of the discussed RG flow of the reduced amplitude. 

It is notable that this analysis rests on using the reduced amplitude, i.e.~where only one term in the exponential expression for EPRL-FK amplitude (the one containing the exponential of the Regge action) is kept. As soon as this is replaced with the full EPRL-FK amplitude, the analysis does not hold any more. This can be traced back to the presence of the cosine, as well as the weird terms. Indeed, in the case where these terms are present, the path integral is a sum over different possibilities, in which different vertices contribute the same parts of the Regge action with different signs. This allows for several terms in which the individual contributions of vertices identically cancel, irrespective of the configuration.  As a result, the stationary phase approximation is dominated by those terms, which do not only contribute the classical solutions, but many non-classical configurations as well. For instance, all transitions via arbitrary intermediate (bulk) spin $j$ contribute. Since the quantum theory is not dominated by the classical solutions in this case, it seems unlikely that the free theory is a fixed point in this case.

Incidentally, the problem, can be avoided when using only the cosine, as well as an odd number of vertices per time step. This is an indication that, for Lorentzian signature and an odd number of vertices, the free theory might indeed be a fixed point. 

\section{Summary and Conclusion}\label{Sec:Summary}

In this article we have investigated the RG flow of the 4d Riemannian EPRL spin foam model for quantum gravity with analytical and numerical tools. For this, several approximations and truncations were employed, in order to make the analysis tractable. 

Previous investigations \cite{Bahr:2016hwc, Bahr:2017klw} only allowed for quasi-local fluctuation of the metric which are, in the semiclassical limit, expected to turn to gauge degrees of freedom. It can be expected that these appear in the theory as spurious degrees of freedom, since it is well-known that the gauge symmetry of GR is broken in the EPRL model \cite{Bahr:2009ku, Bahr:2015gxa, Asante:2018wqy}.

The crucial innovation of this article is to relax previous truncations to include quantum frustal geometries \cite{Bahr:2017eyi}. This allows for curvature fluctuation in the model, which are not just pure gauge. Also, the model restricted to frusta is an extension of the previous setting in \cite{Bahr:2016hwc, Bahr:2017klw}, which allows for degrees of freedom which are local in time. 

The interesting coupling constants of this model are the gravitational and cosmological constants $G$ and $\Lambda$, as well as a parameter in the path integral measure $\alpha$, which is connected to the $4$-volume in the measure, and has been shown to play a crucial role in the restoration of broken diffeomorphism symmetry \cite{Bahr:2016hwc}.

In our analysis, we have worked on hypercubic lattices, which provide discretizations of a torus universe. The RG flow was considered for various coarse graining steps of finer to coarser discretizations. 

To define a flow in terms of coupling constants $G$, $\Lambda$, $\alpha$, it was necessary to choose a couple of reference observables, which we compared on the coarse and the fine lattice. Here, we mostly restricted ourselves to $3$- and $4$-volumes, as well as their fluctuations. Different choices are possible, but we expect those to yield only qualitatively minor changes to the results, as long as one considers obervables which are diverse enough as to separate the space of considered path integral measures. See also discussion in \cite{Bahr:2014qza}.

Furthermore, we employed a system which made the RG flow much more accessible. By relaxing the condition for cylindrical consistency, but allowing only slight changes in the coupling constants, we were able to produce a much smoother flow. As a drawback, the flow diagrams cannot be trusted everywhere, but with the deviation $R$ from cylindrical consistency (\ref{Eq:BIR_CylindricalConsistencyObservables}), we have a control parameter to judge the quality of the resulting flow in any region. This allowed for quick scanning of parts of the phase space, since in the region of fixed points it can be expected that the value of $R$ has to be small. It is in the vicinity of these regions that one can trust the flow images the most.

\subsection{Our findings}

Our results are as follows:

\begin{itemize}
	\item Firstly, the employed approximations allow us to generate images of the RG flow. The introduction of the $R$-parameter allowed us to quickly decide which regions of the phase space are more likely to contain fixed points, and were worthwhile to concentrate our analysis around. This is in general very encouraging, and we believe that this method can also be used more generally in other RG applications, possibly even beyond the spin foam context.

	\item We have considered three main flows. One in the parameter $\alpha$, which was taken to be isochoric, i.e.~with fixed total $4$-volume. This was a direct generalization of the flow computed in \cite{Bahr:2016hwc}, where the non-trivial fixed point was found. Our analysis revealed that the fixed point was still present, albeit with a slightly changed numerical value. We found that in the case of frusta the fixed point lies at 
	\begin{eqnarray}
		\alpha^*\approx 0.69,
	\end{eqnarray}
	
	\noindent which is slightly increased from $\alpha^*\approx 0.63$ in the case of hypercuboids. 
	
	\item For considering the RG flow in more parameters, we considered a $2$d flow in $G$ and $\Lambda$, keeping $\alpha\approx 0.68$ fixed. We used this to scan the phase space for regions likely containing the fixed points, using the procedure describe above. 
	
	We then also considered a $3$d flow in all three parameters $(G,\Lambda, \alpha)$ within that region. We found that there appears to be a fixed point at 
	\begin{eqnarray}
		\alpha^*\approx 0.677, G^*\approx 0.037, \Lambda^*\approx 0.008.
	\end{eqnarray}
	
	\noindent Numerical evidence shows that the fixed point has one repulsive and two attractive directions. 
	
	\item We also considered the \emph{free theory}, i.e.~the point on which the coupling constants $G=0$, $\Lambda=0$. This point plays an important role in the perturbative renormalization of GR, which is defined by perturbations around it. The EPRL model is defined non-perturbatively, which is seen as one of the strengths of the (loop) quantum gravity approach. This, however, makes it difficult to draw comparisons to more traditional forms of the analysis.
	
	In particular, this point is not part of the range of EPRL amplitudes. However, with our methods of defining the flow via observables, we can investigate this point at least asymptotically, since it sits on the infinite boundary of the EPRL theory space, and expectation values of some observables converge when approaching this point. 
	
	In particular, we could approach this point both numerically and analytically by asymptotic methods. We found that, contrary to our assumptions, the free theory appears \emph{not} to be a fixed point of the Riemannian EPRL model. If we replace the EPRL amplitude by the exponential of the Regge action (with measure factors from the asymptotic EPRL amplitude), we however can show that the free theory is a fixed point. 
	
\end{itemize}

\subsection{Discussion}

The main goal of our analysis was to learn more about the RG flow of the EPRL model. Indeed, there are several lessons one might draw from our findings.
\begin{itemize}
	\item The stability of existence of fixed point under extension of the parameters, and relaxing of truncations, fosters hope that this sort of fixed point is an actual feature of the model, rather than an artefact of the approximation. Of course, further study needs to be taken before this point can be settled decidedly. At this instance, it is unclear whether this fixed point is the only interesting one of its kind in the considered phase space. It is also not clear whether this point bears any relation to the non-Gaussian fixed point discussed in the Asymptotic Safety Scenario \cite{Niedermaier:2006wt}.

	\item The fact that the free theory (i.e.~where $G=0$, $\Lambda=0$) is not a fixed point of the EPRL model, but becomes one when replacing it with simply the exponential of the Regge action, was an unexpected feature. It can be understood by the form of the EPRL amplitude: Apart from the exponential of the Regge action, it also contains its sign-reversed part (commonly referred to as the \emph{cosine problem}), as well as other, non-geometric terms (colloquially called \emph{weird terms}). 
	
	It is the presence of these additional terms which spoil the fixed point properties. In the free theory, it should be expected that quantum fluctuations around the classical solution are suppressed, since the pre-factor in front of the Regge action oscillates rapidly for even minor deviations from the classical trajectory. However, in the EPRL amplitude, the situation changes, since terms with opposite signs can cancel each other in the action. Fluctuations in these directions are therefore not suppressed since they do not change the value of the amplitude. These highly curved contributions are quite different numerically on different lattices, which is why the fixed point properties are spoiled. 
	
	The main message one might take away from this is that the Riemannian EPRL model can be expected, in general, to be quite a different theory from (Riemannian signature) quantum gravity. This in itself is not surprising, but, to our knowledge, this is the first instance where this fact has been observed explicitly. It should be noted that in the Lorentzian-signature version of the EPRL model, the weird terms are absent. Also, there is work on the so-called proper vertex, which aims at resolving the cosine issue, even for the Lorentzian amplitude \cite{Engle:2011un, Engle:2015mra}.
	
	The question of whether the two terms in the cosine interfere with one another has not been decisively settled by our analysis, but the question appears to be answered in the affirmative. There are, however, some caveats which might, in the long run, change this point of view:
	
	Firstly, if the weird terms are absent (as happens in the Lorentzian theory), one can make the free theory into a fixed point by only considering lattices with an \emph{odd} number of vertices. This prevents precise cancellation of contributions from vertices with differing signs. Still, this restriction appears slightly artificial to us, but it illustrates an important point: the cancellations also happen because of the large amount of symmetries we consider, i.e.~by using frusta. In the unrestricted  theory where all fluctuations are considered, the states in which precise cancellation among all vertices happens might be dominated by those where it does not. This kind of entropic argument could resolve the issue for the Lorentzian amplitude. 
	
	Secondly, our choice of coherent states might influence the result as well. In general, it is expected that one can restrict to either sign of the action by prescribing the proper extrinsic curvature on the boundary. The Livine-Speziale intertwiners used in our analysis are maximally uncertain in the extrinsic curvature, so that both signs of the Regge action are excited equally. It is feasible to assume that by choosing boundary states which suppresses one sign, one can effectively implement the proper vertex (with minor fluctuations), which would turn the free theory into a fixed point. 
	
	This point certainly warrants further investigations in the future.

\end{itemize}

\acknowledgements

This work was funded in part by the project BA 4966/1-1 of the German Research Foundation (DFG). This research was supported in part by Perimeter Institute for Theoretical Physics. Research at Perimeter Institute is supported by the Government of Canada through Innovation, Science and Economic Development Canada and by the Province of Ontario through the Ministry of Research, Innovation and Science.​ GR would like to thank the research group at CPT, Marseilles, for their hospitality during the finalization of this manuscript. 

\appendix

\bibliography{bibliography}
\bibliographystyle{ieeetr}

\end{document}